\documentclass[journal,twoside]{IEEEtran}
\usepackage{graphicx}
\usepackage{bmpsize}
\usepackage{multicol}
\usepackage{amsmath}
\usepackage{amsfonts}
\usepackage{amsthm}
\usepackage[mathscr]{euscript}
\usepackage{mdwmath}
\usepackage{bbm}
\usepackage{mathrsfs}
\usepackage{bm}
\usepackage{epstopdf}
\usepackage{booktabs}
\usepackage[flushleft]{threeparttable}
\usepackage{siunitx}
\usepackage[tight,footnotesize]{subfigure}
\usepackage{cite}
\usepackage{wrapfig}
\usepackage[justification=centering]{caption}
\usepackage{algorithm}
\usepackage[]{algpseudocode}
\makeatletter
\newcommand{\algmargin}{\the\ALG@thistlm}
\makeatother
\newlength{\whilewidth}
\algdef{SE}[parWHILE]{parWhile}{EndparWhile}[1]
{\parbox[t]{\dimexpr\linewidth-\algmargin}{%
		\hangindent\whilewidth\strut\algorithmicwhile\ #1\ \algorithmicdo\strut}}{\algorithmicend\ \algorithmicwhile}%
\algnewcommand{\parState}[1]{\State%
	\parbox[t]{\dimexpr\linewidth-\algmargin}{\strut #1\strut}}
\theoremstyle{plain}

\usepackage{float}

\usepackage{epstopdf}
\usepackage{comment}

\usepackage{marginnote}
\usepackage{color}

\newcommand{\mydef}{:=}
\newcommand{\dx}{\dot{x}}
\newcommand{\Rc}{\mathbb{R}}

\algnewcommand{\algorithmicgoto}{\textbf{go to}}%
\algnewcommand{\Goto}[1]{\algorithmicgoto~\ref{#1}}%

\allowdisplaybreaks

\begin{document}

\title{Searching for the Shortest Path to the Point of Voltage Collapse on the Algebraic Manifold}
\author{
\IEEEauthorblockN{Dan Wu$^{\dagger}$,
\IEEEmembership{Member, IEEE}, Franz-Erich Wolter$^{\ddagger}$, Bin Wang$^{\star}$, \IEEEmembership{Member, IEEE}, and Le Xie$^{\star}$, \IEEEmembership{Member, IEEE}
}	
\thanks{$\dagger$: Lab for Information and Decision Systems, Massachusetts Institute of Technology, Cambridge, MA, danwumit@mit.edu; $\ddagger$: Department of Electrical Engineering and Computer Science, Leibniz University Hannover, Germany, few@gdv.unihannover.de; $\star$: Department of Electrical and Computer Engineering, Taxes A\&M University, College Station, TX, wangbin.dianqi@gmail.com and le.xie@tamu.edu}%
}
\maketitle
\begin{abstract}
Voltage instability is one of the main causes of power system blackouts. Emerging technologies such as renewable energy integration, distributed energy resources and demand responses may introduce significant uncertainties in analyzing of system-wide voltage stability. This paper starts with summarizing different known voltage instability mechanisms, and then focuses on a class of voltage instability which is induced by the singular surface of the algebraic manifold. We argue and demonstrate that this class can include both dynamic and static voltage instabilities. To determine the minimum distance to the point of voltage collapse, a new formulation is proposed on the algebraic manifold. This formulation is further converted into an optimal control framework for identifying the path with minimum distance on the manifold. Comprehensive numerical studies are conducted on some manifolds of different power system test cases and demonstrate that the proposed method yields candidates for the local shortest paths to the singular surface on the manifold for both the dynamic model and the static model. Simulations show that the proposed method can identify shorter paths on the manifold than the paths associated with the minimum Euclidean distances. Furthermore, the proposed method always locates the right path ending at the correct singular surface which is responsible for the voltage instability; while the Euclidean distance formulation can mistakenly find solutions on the wrong singular surface. A broad range of potential applications using the proposed method are also discussed.
\end{abstract}


\section{Introduction}
   \label{sec:intro}
   Voltage collapse has been widely recognized as a fundamental cause for power system blackouts. Past events include the 1970 New York blackout, the 1987 Tokyo blackout, the 1995 Israel blackout, and the 2003 North America blackout \cite{kaur:2012review,ohno:20061987,hain:1997analysis}. Numerous studies were carried out to analyze\cite{dobson:1989towards,dobson:1992observations,lee:1993dynamic,canizares:1995bifurcations}, monitor\cite{xie:2012distributed,li:2015wide,zheng:2017novel}, mitigate\cite{chevalier:2018mitigating,rabiee:2018optimal}, and control\cite{feng:2000comprehensive,vargas:2000time,urquidez:2015singular,yao:2019optimal} voltage collapse. 
Empirical observations from past events indicate that voltage instability is likely to happen under stressed operating conditions. 
Nowadays the increasing penetration of renewable energy brings more fluctuations and uncertainties in the power grid, introducing more frequent and significant loading condition variations that may cause potential voltage instability. 

A traditional approach which identifies the (quasi) static voltage collapse point evaluates PV-QV curves. It keeps track of the change of a power flow solution at a given direction in the power generation and demand space through the continuation power flow method \cite{iba1991:calculation,chiang:1995cpflow,ghiocel:2013power}. This approach works well with an accurate awareness of the system future operating conditions. However, as renewable energy resources increasingly penetrate into the grid, power generation becomes highly volatile and, thus, can void the voltage stability assessment based on a predicted direction. At transmission level, a prediction error of up to $10\%$ can happen in the day-ahead forecast peak load at ERCOT \cite{ercot:load}. This may easily exhaust the merely available power reserves, which may be lower than $3\%$ in the real time operation \cite{ercot:9000dpmwh}. Without delicate coordinations, these changes can be harmful to voltage stability. When it comes to the distribution network, emerging technologies such as distributed energy resources (DERs) and demand response (DR) can largely alter the properties of traditional load. For example, a distribution system with DERs can inject power to the grid instead of absorbing power. Such power flow reversal provides system operators and customers with much more flexibility but also challenges the conventional control and operation strategies based on traditional load assumptions. A distribution network with DR may change the composition of load in a spatial-temporal manner, resulting in a highly variable aggregated load model. 
These upcoming challenges in both the transmission and the distribution systems should be appropriately addressed for maintaining voltage stability. Therefore, more robust approaches are required to accommodate high dimensional variations of generation and load.

To estimate the voltage stability margin, many indicators and approaches are proposed \cite{kessel:1986estimating,dobson:1991direct,gubina:1995voltage,moghavvemi:1998technique,musirin:2002novel,balamourougan:2004technique,simpson:2016distributed,kamel:2017development}. Comparisons among some of these indicators were discussed in \cite{canizares1996:comparison,chandra:2016comparative}. These indicators, either empirical or analytical, represent different definitions of distances from an operating point to its collapse boundary. 
A common measure of voltage instability proximity in the existing work is the Euclidean distance in the power space. 
An optimization framework was proposed in \cite{jung1990:marginal} to solve the minimum load distance to the singular boundary. A direct approach was proposed in \cite{dobson:1991direct} which is related to the optimization framework in \cite{jung1990:marginal}. Literature \cite{dobson:2003distance} further stated that both direct and indirect methods are approaches to solve the optimization problem. The optimization approach with KKT conditions was further extended in \cite{yao:2018improving}. 

However, as will be shown in this paper, voltage stability measured from the Euclidean distance can be conservative and misleading, especially when the optimization yields the global solution: as shall be shown, multiple isolated singular surfaces can exist but only one contributes to the relevant voltage instability behavior. The existing optimization framework cannot distinguish between different singular surfaces. Furthermore, the actual power trajectory may not follow a straight line because of power flow constraints. Therefore, a more rigorous definition of voltage stability distance is required.

Based on the foundational optimization framework of voltage collapse \cite{jung1990:marginal,dobson:1991direct}, we improve the existing result by searching the shortest path on the algebraic manifold\footnote{The term \emph{algebraic manifold} will be rigorously defined in the next section.} instead of in the Euclidean space. This ``subtle" change requires a completely new formulation of the problem which is fully developed in this paper. Then, a general approach is proposed to solve the problem by an optimal control framework. One should note that the proposed method is not necessarily restricted to the voltage stability problem. It is a general optimization framework that can be used for other problems which seek for shortest paths on manifolds. The major contributions of this paper are summarized below.
\begin{enumerate}
	\item We redefine the shortest distance problem (to the voltage collapse) on the algebraic manifold. It is a new and more rigorous formulation in the sense that it respects physical constraints and voltage behaviors on the manifold for the entire path.
	\item We convert the voltage stability problem into an optimal control problem and solve it to acquire the shortest path on the manifold to the point of voltage collapse. 
	\item We compare the results of the proposed formulation with the results of the Euclidean distance formulation, explain the conservativeness and misleading caveats of the Euclidean distance in some cases, and show the correctness of proposed formulation. 
\end{enumerate}

The rest of the paper is organized as follow. Section~\ref{sec:mechanism} discusses the known mechanisms of voltage collapse and defines a class of voltage stability problem that will be investigated in this paper. Section~\ref{sec:traditional} introduces the optimization framework based on the Euclidean distance. Section~\ref{sec:manifold} proposes a new formulation based on the manifold distance and reformulates it as an optimal control problem. Numerical simulations are provided in Section~\ref{sec:num}. Section~\ref{sec:disc} provides some discussions and remarks on the technical details and potential applications. Setion~\ref{sec:concl} concludes the paper.
   
\section{Mechanisms of Voltage Collapse}
   \label{sec:mechanism}
Based on different features and time scales of instability, some researchers classify voltage stability analysis into two categories: dynamic and static (or quasi-static) \cite{kundur1994:power}. The static voltage instability is assumed to happen when a system is perturbed in its parameter space and reaches a saddle-node bifurcation point\footnote{Reaching an engineering limit can also induce voltage instability, for example, the generator reactive power limit. But this is beyond the scope of this paper.}. The saddle-node bifurcation is determined by the singular surface\footnote{The word ``surface'' in this paper does not necessarily imply a 2-D surface. Here we refer to a general multi-dimensional hyper-surface.} of algebraic equations. During the perturbation process, the system is assumed to be capable of stabilizing at the newly perturbed equilibrium point until reaching singularity. 

On the other hand, the dynamic voltage instability usually happens during the transient process in which the system states are away from a stable equilibrium point (SEP). The cause of dynamic voltage instability can be non-unique. 
For example, \cite{chiang:1990voltage,van2007:voltage} discussed the influence of center manifolds in voltage collapse models. \cite{wu:2006geometrical} observed that some unstable equilibrium points (UEP) are responsible for voltage instability.  \cite{oluic2018:nature,wu:2019influence} showed that singular surface of algebraic constraints can also contribute to the dynamic voltage instability.

Below we recall both dynamic and static voltage instability, attempt to unify different models into the dynamic-algebraic equation form, and focus on a class of voltage instability, including the above-mentioned static and dynamic ones.

\subsection{Dynamic Voltage Instability}
A power system dynamic model usually takes the differential-algebraic equation (DAE) form
\begin{subequations}
	\begin{eqnarray}
	\dx &=& f(x,y) \label{eq:DAE_D}\\
	0 &=& g(x,y) \label{eq:DAE_A}
	\end{eqnarray} \label{eq:DAE}%
\end{subequations}
where $x\in \Rc^m$ includes dynamic states, e.g., rotor angles and angular velocities in swing equations\footnote{Detailed dynamic models can introduce more dynamic states such as field winding voltages, AVR internal voltages, etc.}; $\dx$ is the time derivative of $x$; $y \in \Rc^n$ includes algebraic states; $f:\Rc^{m+n} \to \Rc^m$ and $g:\Rc^{m+n} \to \Rc^n$ are continuously differentiable. 

The stability boundary of the DAE system \eqref{eq:DAE} is determined by both the dynamic part and the algebraic part \cite{venkatasubramanian:1992stability,venkatasubramanian:1995dynamics,chiang:2015stability}. 
The dynamic part is related to the type-1 unstable equilibrium points (UEPs). To identify the type of an equilibrium, consider the linear form of \eqref{eq:DAE} at the equilibrium.
\begin{subequations}
	\begin{eqnarray}
	\delta \dx &=& \frac{\partial f}{\partial x} \delta x + \frac{\partial f}{\partial y} \delta y \label{eq:JDAE_D}\\
	0 &=& \frac{\partial g}{\partial x} \delta x + \frac{\partial g}{\partial y} \delta y \label{eq:JDAE_A}
	\end{eqnarray} \label{eq:JDAE}%
\end{subequations}
It can be further reduced to
\begin{equation}
	\delta \dx = \bigg(\frac{\partial f}{\partial x} - \frac{\partial f}{\partial y}~ \big(\frac{\partial g}{\partial y}\big)^{-1}~ \frac{\partial g}{\partial x} \bigg) \delta x \label{eq:JODE}
\end{equation}
provided $\partial g / \partial y$ is invertible. The type of an equilibrium point is determined by the eigenvalues of the Jacobian matrix in \eqref{eq:JODE}. A hyperbolic equilibrium point that has only one eigenvalue with positive real part is said to be type-1.

On the other hand, the singular surface\footnote{The term ``singular surface'' will be defined shortly below.} of algebraic constraints can also contribute to the dynamic instability phenomenon \cite{venkatasubramanian:1992stability,venkatasubramanian:1995dynamics,chiang:2015stability}. A comprehensive numerical study of singularity induced dynamic voltage instability can be found in \cite{oluic2018:nature}, which is also confirmed in \cite{wu:2019influence}.

According to the analysis in \cite{lee:1993dynamic,van2007:voltage,wu:2006geometrical,chiang:1990voltage,oluic2018:nature,wu:2019influence},  dynamic voltage instability can happen if 
\begin{enumerate}
	\item Eqt~\eqref{eq:JODE} is singular at the equilibrium point (central manifold occurs);
	\item a trajectory is beyond the stable manifolds of some type-1 UEP;
	\item a trajectory crosses the algebraic singular part of the stability boundary.
\end{enumerate}
\begin{figure}[tb!]
	\centerline{\includegraphics[width=0.8\columnwidth]{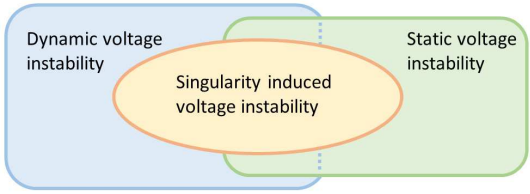}}
	\caption{Voltage Instability in Different Time Scales}
	\label{fig:vc}
\end{figure}

\subsection{Static Voltage Instability}
	Static voltage instability is usually characterized by the singularity condition of the power flow Jacobian. When changing some nodal power injections, two real-valued power flow solutions collide with each other and become a pair of complex-valued conjugate solutions that are not physically realistic.
	This phenomenon seems to be purely algebraic since it only characterizes how a set of algebraic equations lose a pair of real-valued solutions. 
	Neither dynamic states nor dynamic equations are specified. 
	
	According to \cite{van2007:voltage}, this power flow based algebraic model is called the \emph{network-only model}. It is simplified from the \emph{quasi-steady state approximation of long-term dynamics} by eliminating all the long term dynamic states and their equations. The general model of quasi-steady state approximation of long-term dynamics takes the following form.

\begin{subequations}
	\begin{eqnarray}
	x(t) &=& \Phi \big(x(t_0),y(t_0), t_0, t \big) \label{eq:DyAE_Dy}\\
	0 &=& g(x,y) \label{eq:DyAE_A}
	\end{eqnarray} \label{eq:DyAE}%
\end{subequations}
where $x(t)$ includes the long term dynamic states. They can include, but are not limited to, load response, renewable fluctuations, secondary controls, distributed energy resources, generation re-dispatch, etc. In our problem formulation, we put all the time varying or adjustable nodal power injections (generation and load) in the long term dynamic state vector $x(t)$. They are traditionally treated as parameters in power system literature.
The algebraic state vector $y(t)$ includes the network node voltage magnitudes and angles in the power flow equations. $y(t)$ does not have independent dynamics. Instead, it changes with respect to time $t$ according to the change of $x(t)$ to satisfy the algebraic constraints \eqref{eq:DyAE_A}. $x(t_0)$ and $y(t_0)$ are the initial values. 
For the dynamic equation \eqref{eq:DyAE_Dy}, we only require $\Phi$ to be a well-defined and slow\footnote{``Slow'' means the dynamics behave several orders of magnitude slower than the ignored fast dynamics. This assumption is valid when traditional load demand and generation vary in minutes while controls and electro-magnetics change in dozens of milliseconds. However, with more power electronics interface installed, the traditional static assumption may require a revisit.} time-forwarding iterative process. $\Phi$ can take the form of differential or integral equations, discrete time iterations, random process, or their hybrid. In power systems, such slow dynamics can include, but are not limited to, load demand change, generator re-dispatch, renewable fluctuation, transformer tap ratio change, automatic generation control, inter-area frequency control, etc. Similar to the DAE system in \eqref{eq:DAE}, Eqt~\eqref{eq:DyAE} is also an algebraically constrained dynamical system for which the stability boundary comprises both the dynamic part and the algebraic part. 
The algebraic part is described by the singularity condition of $\partial g / \partial y$, which is the singular Jacobian matrix of power flow equations \cite{venkatasubramanian:1992stability,venkatasubramanian:1995dynamics}.

\subsection{Voltage Instability Induced by Singular Surface of Algebraic Manifold}
From the above discussions we have shown that algebraic constraints contribute to both dynamic and static voltage instabilities. We define the $m$-dimensional \emph{algebraic manifold}~\footnote{This terminology is used to distinguish from the stable and unstable manifolds of an equilibrium in the dynamic sense.} $\Sigma$ as
\begin{equation}
	\Sigma \mydef \{x \in \Rc^m, y \in \Rc^n~|~g(x,y)=0\}. \label{eq:algebraic_manifold}%
\end{equation} 

A dynamic-algebraic system, either in the form of \eqref{eq:DAE} or \eqref{eq:DyAE}, is a constrained dynamic system whose dynamic flow is confined on $\Sigma$. The singular surface of $\Sigma$, denoted by $\Omega$, 
\begin{equation}
	\Omega \mydef \{x \in \Rc^m, y \in \Rc^n~|~g(x,y)=0,~\text{det}(\frac{\partial g}{\partial y})=0\} \label{eq:algebraic_singular}%
\end{equation}
can induce a certain class of voltage instability which can be either dynamic or static. Therefore in our discussions, we do not distinguish if a voltage instability is dynamic or static, but only focus on the singular surface $\Omega$ induced voltage problem (as shown in Fig.~\ref{fig:vc}). Our goal is to identify the shortest path to the singular boundary $\Omega$ on the manifold $\Sigma$. 

\section{Formulation Based on Euclidean Distance}
	\label{sec:traditional}
	We start our discussion with the existing problem formulation in the Euclidean space. Proposed in \cite{jung1990:marginal} and extended in \cite{dobson:1991direct,yao:2018improving}, the formulation in Euclidean distance is built rigorously on an optimization framework and holds an alternative simple representation for the singularity condition of \eqref{eq:algebraic_singular}. A general formulation of this kind is
\begin{subequations}
	\begin{eqnarray}
	\text{min:} & &|z_c^0-z_c|_2^2 \label{eq:Euclidean_cost}\\
	\text{s.t.:} & &g(x,y)=0 \label{eq:Euclidean_alg}\\
	 & &\big( \partial g / \partial y \big)^T r = 0 \label{eq:Euclidean_sing}\\
	 & &\langle r, r \rangle =1 \label{eq:Euclidean_unif}
	\end{eqnarray} \label{eq:Euclidean}%
\end{subequations}
where based on our definition in \eqref{eq:DyAE} $x \in \Rc^m$ includes the dynamical states which are time varying or adjustable parameters in the traditional sense; $y \in \Rc^n$ includes the algebraic states; $z_c \subseteq (x,y)$ is a subset of states $x$ and $y$; $z_c^0$ is a given constant vector representing the known operating point; $r \in \Rc^n$ is the vector of auxiliary states for enforcing the singularity condition; $\langle a,b \rangle$ is the inner product operator of vector space. 

Usually $z_c$ is selected from the generator power output and load demand. But there is no limit to selecting other state variables. For example, we can also include bus voltages in $z_c$. The goal of \eqref{eq:Euclidean} is to find a point $(x^\star,y^\star,r^\star)$ in the augmented state space such that \eqref{eq:Euclidean_alg}, \eqref{eq:Euclidean_sing}, and \eqref{eq:Euclidean_unif} are satisfied and the Euclidean distance between $z_c^0$ and $z_c^\star$ is the minimum (at least locally). 

Eqt. \eqref{eq:Euclidean_alg} enforces that the optimum solution must be on the algebraic manifold $\Sigma$. \eqref{eq:Euclidean_sing} utilizes the auxiliary states $r$ to ensure that the optimum solution is on the singular surface $\Omega$. Vector $r$ can be regarded as a left eigenvector associated with the zero eigenvalue of $\partial g / \partial y$. To acquire a unique $r$, \eqref{eq:Euclidean_unif} is applied to restrict the radius of it. Other constraints can also be included in \eqref{eq:Euclidean}, for example, state variable upper and lower bounds. For simplicity, we only consider indispensable constraints that are listed in \eqref{eq:Euclidean}.

\section{Formulation Based on Manifold Distance}
	\label{sec:manifold}
	The previous section introduced the optimization formulation based on the Euclidean distance. This formulation only ensures that the end point is on the manifold, ignoring the entire transition from the known operating point to the end point. In this section, we will include constraints that force the entire path on the manifold, develop a new optimization framework, and convert it into an optimal control problem. 

\subsection{A General Formulation Based on Manifold Distance}
To establish the distance on the algebraic manifold, the following general formulation is considered. 
\begin{subequations}
	\begin{eqnarray}
	\text{min:} & &\text{arclength}\big( \text{path}(z_c^0,z_c^L) \big) \label{eq:manifold_cost}\\
	\text{s.t.:} & &g\big( \text{path}(x^0,x^L),\text{path}(y^0,y^L) \big)=0 \label{eq:manifold_alg}\\
	& &\big( \partial g / \partial y \big)|_{(x^L,y^L)}^T r^L = 0 \label{eq:manifold_sing}\\
	& &\langle r^L, r^L \rangle =1 \label{eq:manifold_unif}
	\end{eqnarray} \label{eq:manifold}%
\end{subequations}
where path($a,b$) defines a continuous path from point-$a$ to point-$b$; arclength($l$) is the arc length operator that we use for the distance on the manifold.

Constraint \eqref{eq:manifold_alg} requires that any point on the path from $(x^0,y^0)$ to $(x^L,y^L)$ must be on the manifold $\Sigma$. Eqt.~\eqref{eq:manifold_sing} certifies that the ending point $(x^L,y^L)$ of the path is on the singular surface $\Omega$. \eqref{eq:manifold_unif} ensures a (locally) unique ending point of auxiliary state $r$. 
The goal of \eqref{eq:manifold} is to find a path which starts from a known point, ends on the singular surface, and is confined on the manifold such that its arc length is the minimum.

\begin{figure}[tb!]
	\centerline{\includegraphics[width=0.5\columnwidth]{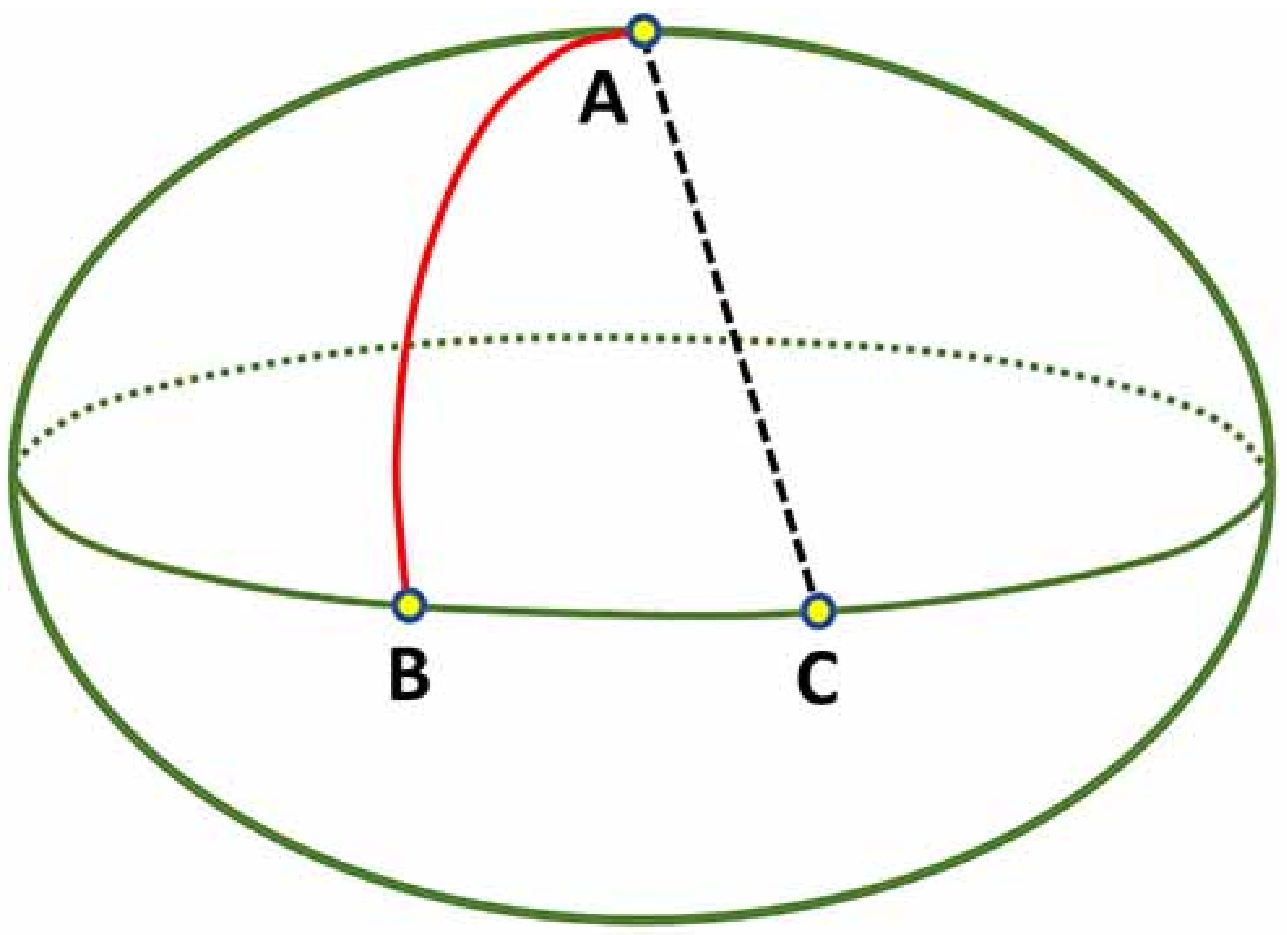}}
	\caption{Path on Manifold VS Euclidean Line Segment}\label{fig:cartoon}
	{\small Red curve: path on manifold. Black dash line: Euclidean line.}
\end{figure}
Fig.~\ref{fig:cartoon} shows two ways to travel from the north pole to the equator. One can dig a straight hole (black dash line segment) through the earth from point-A to point-C, which corresponds to the solution of \eqref{eq:Euclidean}, or take an airplane following the surface of the earth (red curve) from point-A to point-B, which corresponds to the solution of \eqref{eq:manifold}. 

\subsection{Reformulation in Optimal Control Framework}
To solve the optimal path of \eqref{eq:manifold}, one needs a well-defined mathematical formulation of path($a,b$) and arclength($l$). A straightforward way is to parameterize path($a,b$) by a free variable $\tau$ in a given interval, say, $[0,1]$. If we further assume that path($a,b;\tau$) is almost everywhere continuously differentiable\footnote{It means that the measure of discontinuity of the derivative with respect to $\tau$ is zero.} with respect to $\tau$, then \eqref{eq:manifold} can be reformulated in the following way.
\begin{subequations}
	\begin{eqnarray}
	\text{min:} & &\int_0^1 \sqrt{\langle dz_c(\tau)/d\tau,dz_c(\tau)/d\tau \rangle} d\tau \label{eq:para_cost}\\
	\text{s.t.:} & &g\big( x(\tau),y(\tau) \big)=0 \label{eq:para_alg}\\
	& &\big( \partial g / \partial y \big)|_{\tau=1}^T r(1) = 0 \label{eq:para_sing}\\
	& &\langle r(1), r(1) \rangle =1 \label{eq:para_unif}\\
	& &x(0)=x^0 \label{eq:para_ini_x}\\
	& &y(0)=y^0 \label{eq:para_ini_y}
	\end{eqnarray} \label{eq:para}%
\end{subequations}
where the arclength\footnote{This arclength is the integral of the path's velocity vector. The length is measured via the Riemannian metric on manifold defined by \eqref{eq:para_alg}.} is computed by the integral of the path directional derivatives in \eqref{eq:para_cost}; \eqref{eq:para_alg} ensures that the entire path $(x(\tau),y(\tau))$ is on the manifold $\Sigma$; \eqref{eq:para_sing} is the singularity condition for the final state at $\tau=1$; \eqref{eq:para_ini_x} and \eqref{eq:para_ini_y} enforce a given initial state for the path.

To solve \eqref{eq:para}, we introduce new variables $u(\tau)$ and $v(\tau)$, then convert \eqref{eq:para} into an optimal control problem in \eqref{eq:oc}.
\begin{subequations}
	\begin{eqnarray}
	\text{min:} & &\int_0^1 \sqrt{\langle \zeta_c(\tau),\zeta_c(\tau) \rangle} d\tau \label{eq:oc_cost}\\
	\text{s.t.:} & &dx(\tau) /d \tau = u(\tau) \label{eq:oc_diff_x}\\
	& &dy(\tau)/d \tau = v(\tau) \label{eq:oc_diff_y}\\
	& &dr(\tau) /d \tau = 0 \label{eq:oc_diff_r}\\
	& &g\big( x(\tau),y(\tau) \big)=0 \label{eq:oc_alg}\\
	& &\big( \partial g / \partial y \big)|_{\tau=1}^T r(1) = 0 \label{eq:oc_sing}\\
	& &\langle r(1), r(1) \rangle =1 \label{eq:oc_unif}\\
	& &x(0)=x^0 \label{eq:oc_ini_x}\\
	& &y(0)=y^0 \label{eq:oc_ini_y}
	\end{eqnarray} \label{eq:oc}%
\end{subequations}
where $u(\tau)$ and $v(\tau)$ are the control variables associated with the state variables $x(\tau)$ and $y(\tau)$; $\zeta_c(\tau) \subseteq \big( u(\tau),v(\tau) \big)$ are associated with the state variable $z_c(\tau)$; Eqt \eqref{eq:oc_diff_x}, \eqref{eq:oc_diff_y} and \eqref{eq:oc_diff_r} relate the state derivatives to the control variables\footnote{The auxiliary state $r(\tau)$ can have an arbitrary dynamic behavior as long as its final state value can reach every point on the unit ball of \eqref{eq:oc_unif}. For simplicity, we define $r(\tau)$ to be constant with zero derivative in \eqref{eq:oc_diff_r}.} and serve as the dynamic part of the optimal control problem; \eqref{eq:oc_alg} is the path constraint that confines the path on the manifold $\Sigma$; \eqref{eq:oc_sing} and \eqref{eq:oc_unif} are the final state boundary conditions in which \eqref{eq:oc_sing} enforces the final state on the singular surface $\Omega$; \eqref{eq:oc_ini_x} and \eqref{eq:oc_ini_y} are the initial state boundary conditions. We only assume that $u(\tau)$ and $v(\tau)$ are almost everywhere continuous, or in the Lebesgue integrable space. A detailed example of \eqref{eq:oc} for the static voltage stability problem is presented in Appendix~\ref{sec:Appendix II}.

The goal of \eqref{eq:oc} is to find a control trajectory $u(\tau)$ and $v(\tau)$ on $[0,1]$ that steers the states $x(\tau)$ and $y(\tau)$ from a given point $(x^0,y^0)$ to the singular surface $\Omega$ such that the entire path resides on the manifold $\Sigma$ and the arc length is the minimum (at least locally).
   	
\section{Numerical Simulations}
   	\label{sec:num}
   	Numerical simulations are conducted in Matlab 2017 environment on a 64-bit personal computer with an Intel i7 2.8GHz CPU and 16GB RAM. The primal dual interior point solver ``IPOPT''\cite{wachter2006:implementation} is used for solving the Euclidean formulation \eqref{eq:Euclidean}. The optimal control solver ``ICLOCS2'' \cite{nie:2018iclocs2,nie2019:efficient,nie2019:external} is used for solving the manifold formulation \eqref{eq:oc}. 

\subsection{9-Bus Dynamic Voltage Instability Example - Distance in Active Power Subspace}
\begin{figure}[tb!]
	\centerline{\includegraphics[width=0.6\columnwidth]{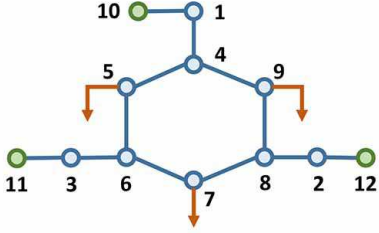}}
	\caption{9-Bus System with Generator Internal Bus}
	\label{fig:9-bus}
\end{figure}

Section~\ref{sec:mechanism} established the focus of this paper on a class of voltage instability which is induced by the singular surface of the algebraic manifold. This class of instability can be either dynamic or static. Our first example is a dynamic one. 

Consider the modified 9-bus example, ``case9mod1'', in \cite{wu:2019influence}. A one-line diagram is depicted in Figure~\ref{fig:9-bus}. 
The transmission system from node-$1$ to node-$9$ in Figure~\ref{fig:9-bus} is unchanged as the IEEE 9-bus system. 
Each original PV bus is modified to a transit PQ bus\footnote{A PQ bus with zero power injection.} connected to a new internal generator PV bus, numbered from node-$10$ to node-$12$ in Figure~\ref{fig:9-bus}. The parameters for this system is given in Table~\ref{table:9bus_1_bus} and \ref{table:9bus_1_branch} in Appendix~\ref{sec:Appendix I}.
We treat $40\%$ of the load demand in Table~\ref{table:9bus_1_bus} as constant impedance and leave the rest $60 \%$ as constant power. The overall system equations are in the DAE form of \eqref{eq:DAE}.

By choosing the center-of-inertial (COI) reference frame, \cite{wu:2019influence} showed that the dynamic stability boundary of this example coincides with the singular surface, suggesting that any transient instability is caused by the singularity. 

The dynamic states of this example are chosen to include the generator relative angles and angular velocities.\footnote{Only two generators are independent.} 
Thus, in the power flow equations $g(x,y)=0$ the generator angles are included in the dynamic states $x$ but not in the algebraic states $y$. It implies that the singularity condition should be applied to the reduced Jacobian matrix $\partial g/\partial y$ instead of the full Jacobian matrix of the power flow problem. 
In this particular example, we consider the minimum distance in the generator active power subspace. Later in the next a few examples, we will consider the minimum distance in the generator reactive power subspace, and a combination of generator and load complex power subspace. 

\begin{figure}[tb!]
	\begin{center}
		\subfigure[Minimum Euclidean Distance{\textsuperscript{a}}]{\label{fig:9_V_P_Euc}\includegraphics[width=0.48\columnwidth]{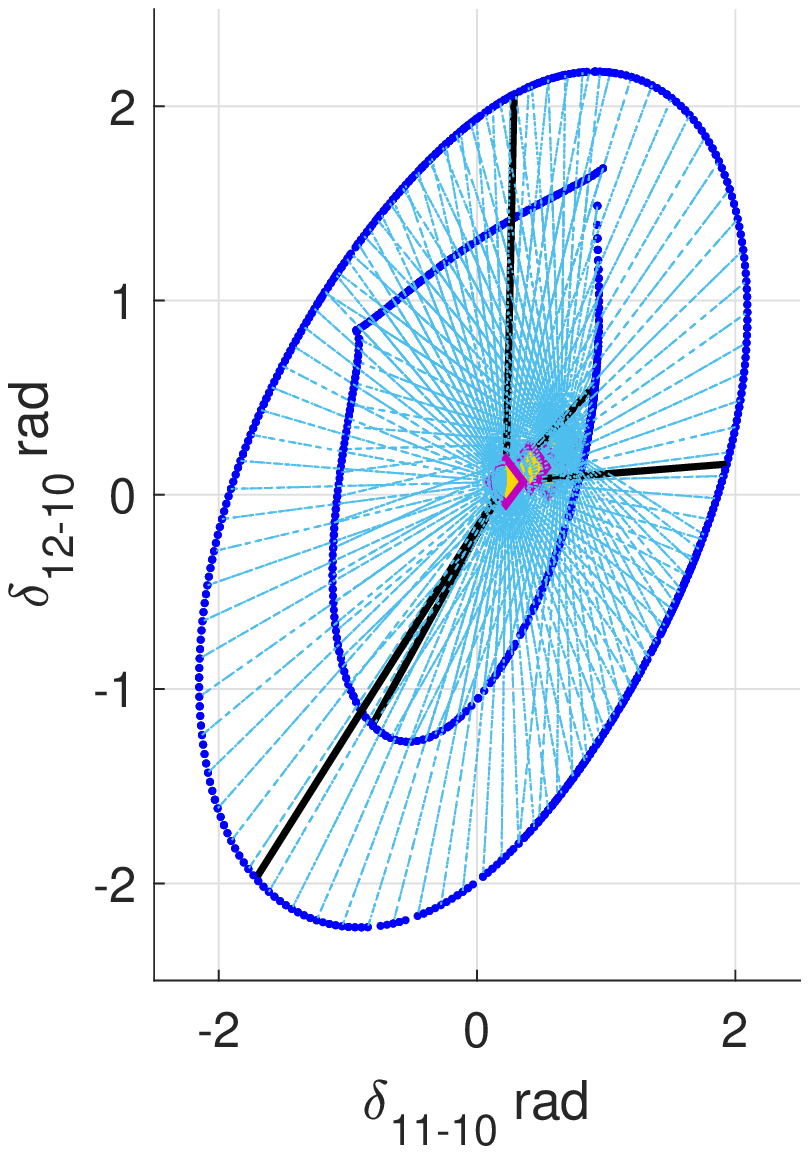}}
		\subfigure[Minimum Manifold Distance{\textsuperscript{b}}]{\label{fig:9_V_P_mnfld}\includegraphics[width=0.48\columnwidth]{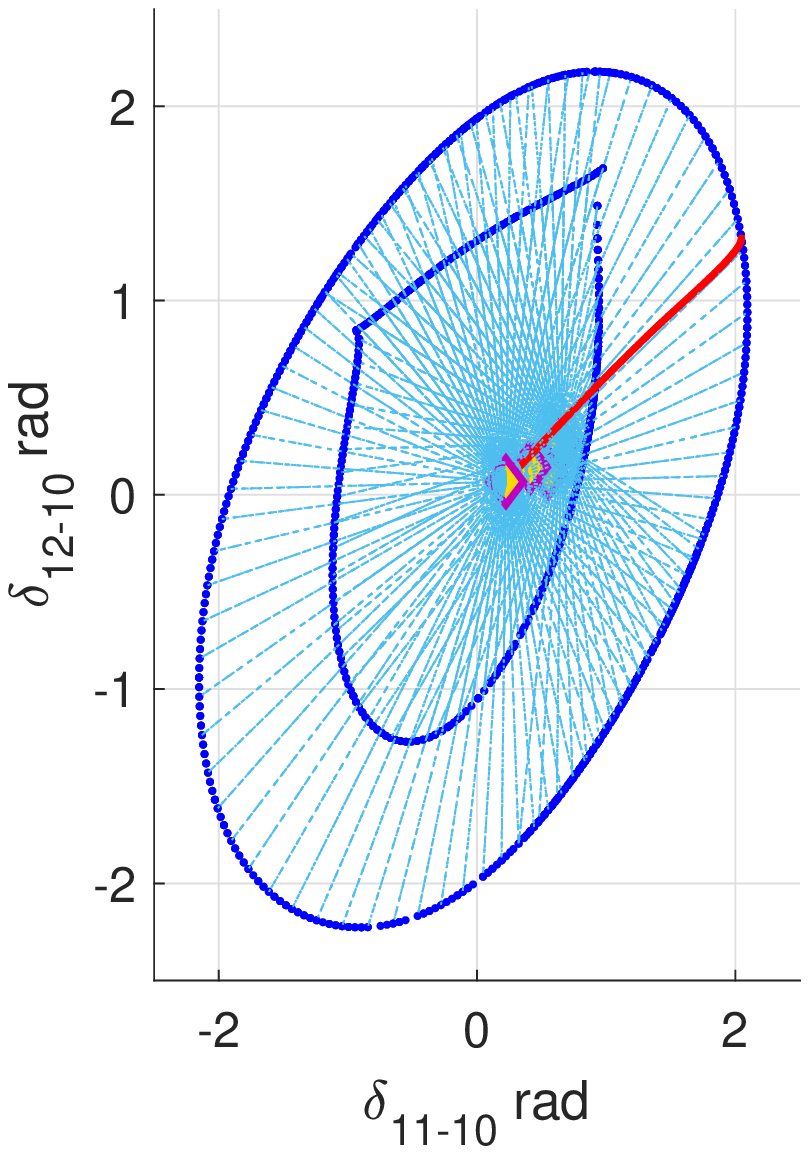}}\\
		\subfigure[Path of Euclidean Distance VS Path of Manifold Distance{\textsuperscript{c}}]{\label{fig:9_V_P_comp}\includegraphics[width=0.85\columnwidth]{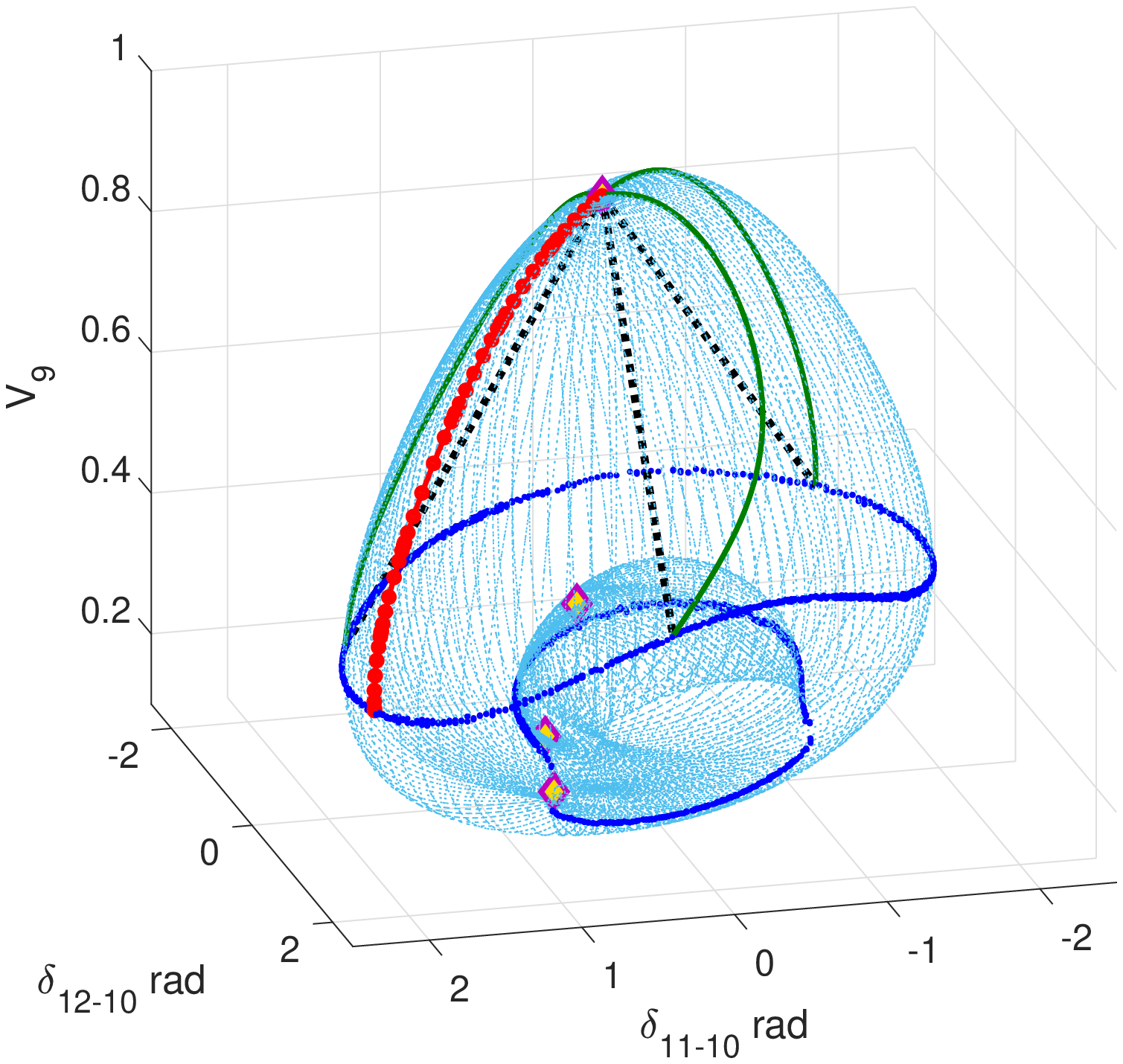}}
		\caption[Caption for LOF]{Projections in Angle-Voltage Subspace for Case9mod1 Dynamic Case}
		\label{fig:9-bus_V_distP_EM}
	\end{center}
	{\small\textsuperscript{a,b,c}Blue dotted curve: singular surface. \small\textsuperscript{a}Black line segment: Euclidean local min. \small\textsuperscript{b}Red curve: manifold local min. 
	\small\textsuperscript{c}Black dotted line segment: Euclidean local min, green curve: associated path for Euclidean local min, red dotted curve: manifold local min, yellow diamond: equilibrium. }
\end{figure}

\begin{figure}[tb!]
	\begin{center}
		\subfigure[Minimum Euclidean Distance{\textsuperscript{a}}]{\label{fig:9_P_P_Euc}\includegraphics[width=0.48\columnwidth]{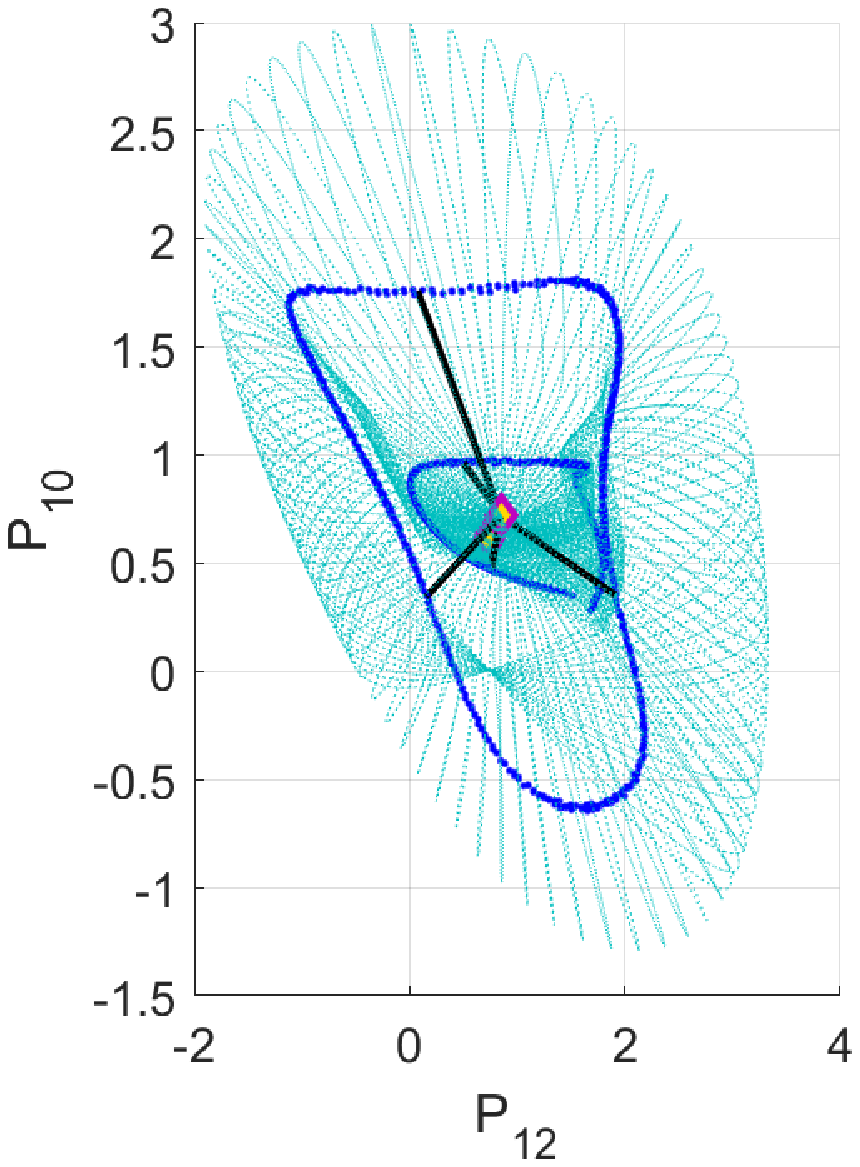}}
		\subfigure[Minimum Manifold Distance{\textsuperscript{b}}]{\label{fig:9_P_P_mnfld}\includegraphics[width=0.48\columnwidth]{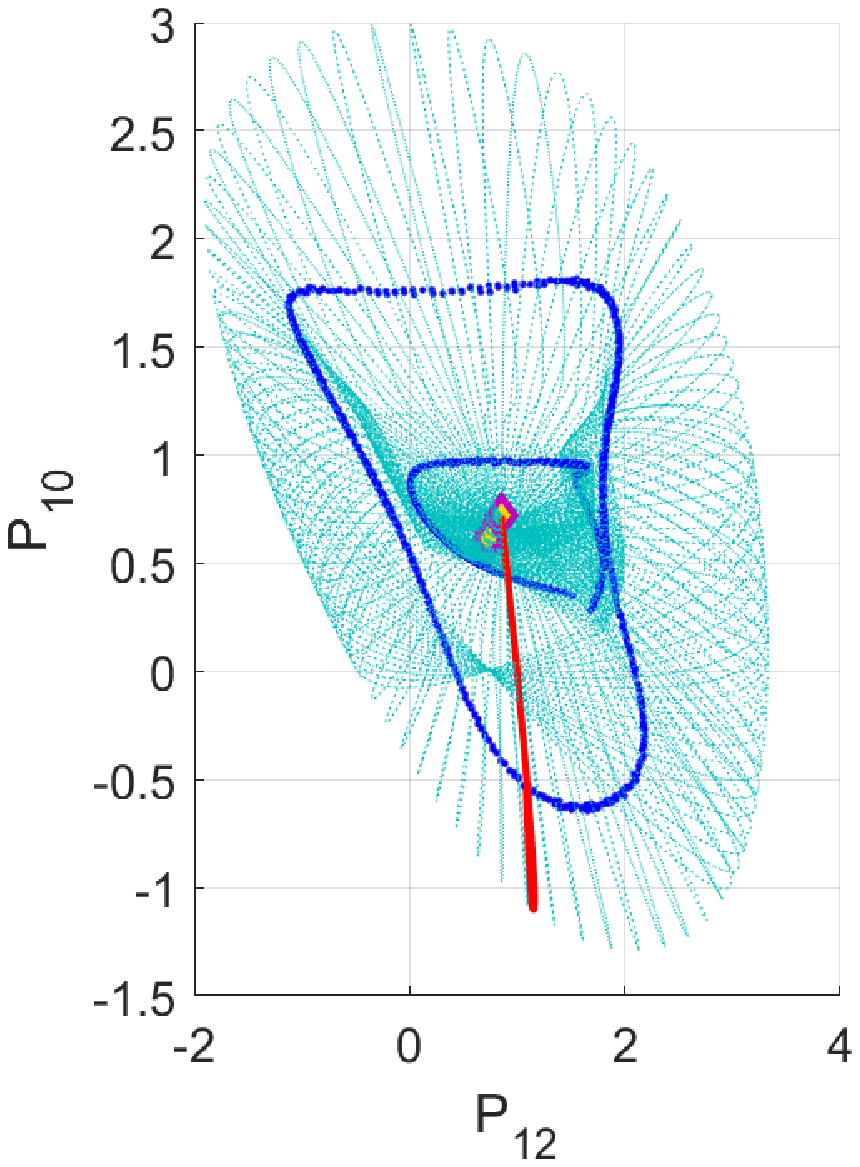}}\\
		\subfigure[Path of Euclidean Distance VS Path of Manifold Distance{\textsuperscript{c}}]{\label{fig:9_P_P_comp}\includegraphics[width=0.85\columnwidth]{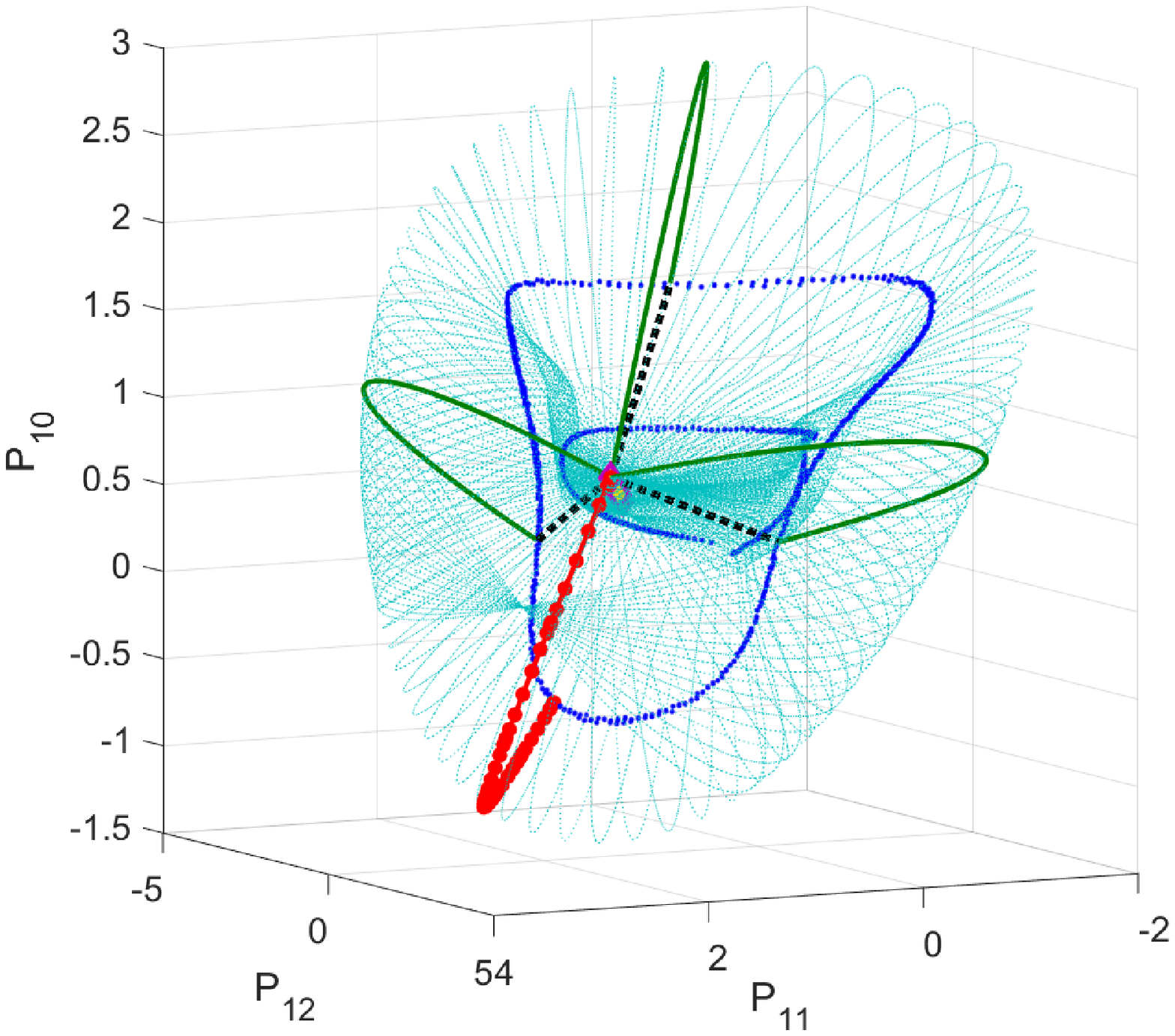}}
		\caption{Projections in Active Power Subspace for Case9mod1 Dynamic Case}
		\label{fig:9-bus_P_distP_EM}
	\end{center}
	{\small\textsuperscript{a,b,c}Blue dotted curve: singular surface. \small\textsuperscript{a}Black line segment: Euclidean local min. \small\textsuperscript{b}Red curve: manifold local min. 
	\small\textsuperscript{c}Black dotted line segment: Euclidean local min, green curve: associated path for Euclidean local min, red dotted curve: manifold local min, yellow diamond: equilibrium.}
\end{figure}

Figure~\ref{fig:9_V_P_comp} depicts a projection of the 2-D algebraic manifold (light blue surface) in a 3-D Angle-Voltage subspace. A local area is blown up inside the manifold creating an inner bubble. This manifold has two isolated singular surfaces (deep blue curves) each of which forms a loop. The lower and smaller one is created by the inner bubble. Four equilibrium points are identified on this manifold (shown in yellow diamonds), in which the upper one is the stable high voltage solution. Figure~\ref{fig:9_P_P_comp} shows the same manifold in another projection which is in the generator active power subspace.  

Firstly, we demonstrate that the results on Euclidean distance can be misleading and conservative. Figure~\ref{fig:9_V_P_Euc} and \ref{fig:9_P_P_Euc} present five candidates\footnote{They are candidates of local minimum because the primal-dual interior point method only solves the first order necessarily optimality condition.} of local minimum (black line segments) found with the Euclidean distance in \eqref{eq:Euclidean}. Three of them end on the larger singular surface, while the other two end on the smaller singular surface. Since the distance is defined in the generator active power space, Figure~\ref{fig:9_P_P_Euc} and numerical results suggest that the global optimum\footnote{We are confident about the global optimum because in this example we can visualize the full manifold and attempt to exhaust all the possible local minima.} resides on the smaller singular surface, i.e., the short black line segment in the 6:30 o'clock direction in Figure~\ref{fig:9_P_P_Euc}. However, this global solution is both misleading and conservative if it is used to determine the voltage stability margin. It is misleading because before reaching the smaller singular surface the system is already destabilized by the larger singular surface. This can be seen from Fig.~\ref{fig:9_V_P_comp} since the inner bubble is created on the lower part of the outer surface. To reach the lower part, one must first cross the larger singular surface. Moreover, following the Euclidean global minimum on the manifold does not necessarily yield a path towards the inner bubble. In Fig.~\ref{fig:9_V_P_comp} one can see that the inner bubble is reachable only for a small range of voltage angle directions. In this particular example, the Euclidean global minimum does not lie in this accessible range. On the other hand, the smaller singular surface encloses a quite smaller region in the generator power space in Figure~\ref{fig:9_P_P_Euc}, which is much smaller than the actual stability boundary given by the larger singular surface. Thus, the Euclidean global solution is also conservative.

Secondly, we argue that our proposed formulation is more meaningful at the global solution. Figure~\ref{fig:9_V_P_mnfld} and \ref{fig:9_P_P_mnfld} present the only minimum distance (red curve) solved on the manifold for \eqref{eq:oc}. These figures clearly suggest that the global solution to \eqref{eq:oc} must end on the larger singular surface because any path that ends on the smaller singular surface must cross the larger one first. It cannot be shorter than a coincident path which only ends on the larger singular surface. However, as the Euclidean distance disrespects the manifold curvature, it may yield the global solution which resides on the smaller singular surface. Thus, the global  intrinsic minimum distance with respect to the interior geometry of the algebraic manifold is more meaningful than the global minimum distance of the Euclidean distance.

Finally, we illustrate that the actual trajectory associated with an Euclidean distance is not a local minimum path on the manifold. In Figure~\ref{fig:9_V_P_comp} and \ref{fig:9_P_P_comp}, we depict three local minima with Euclidean distance (black dot line segments) which end on the correct singular surface and the minimum path on the manifold (red dot curve). Recall that the system trajectory must be on the manifold. If we force the system to follow the directional change of the selected Euclidean distance line segments, the corresponding paths on the manifold can be determined in our particular examples (shown in deep green curves).
We only show the paths associated with the three Euclidean minima ending at the larger singular surface because they are the ones that reach the correct singular surface.
Numerical calculations show that the manifold distance for the red curve is $3.1468$, while the manifold distance for the green curves are $4.3403$, $5.0978$, and $5.7572$, respectively. Hence, the length of the correct path associated with the Euclidean distance at least exceeds the length of the shortest path on the manifold by $37.9\%$ for this particular example. 

\subsection{9-Bus Static Voltage Instability Example - Distance in Active Power Subspace}
\begin{figure}[tb!]
	\begin{center}
		\subfigure[Minimum Euclidean Distance{\textsuperscript{a}}]{\label{fig:9_V_P_full_Euc}\includegraphics[width=0.48\columnwidth]{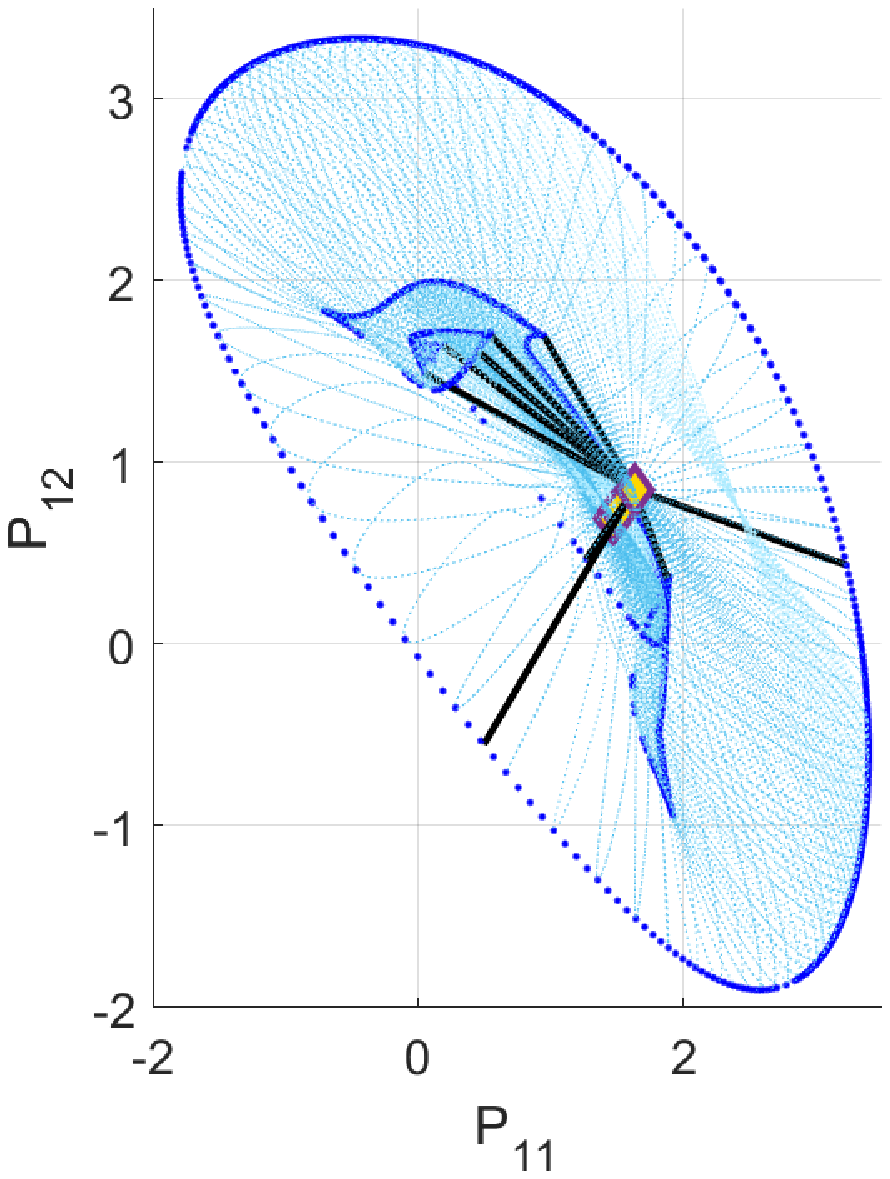}}
		\subfigure[Minimum Manifold Distance{\textsuperscript{b}}]{\label{fig:9_V_P_full_mnfld}\includegraphics[width=0.48\columnwidth]{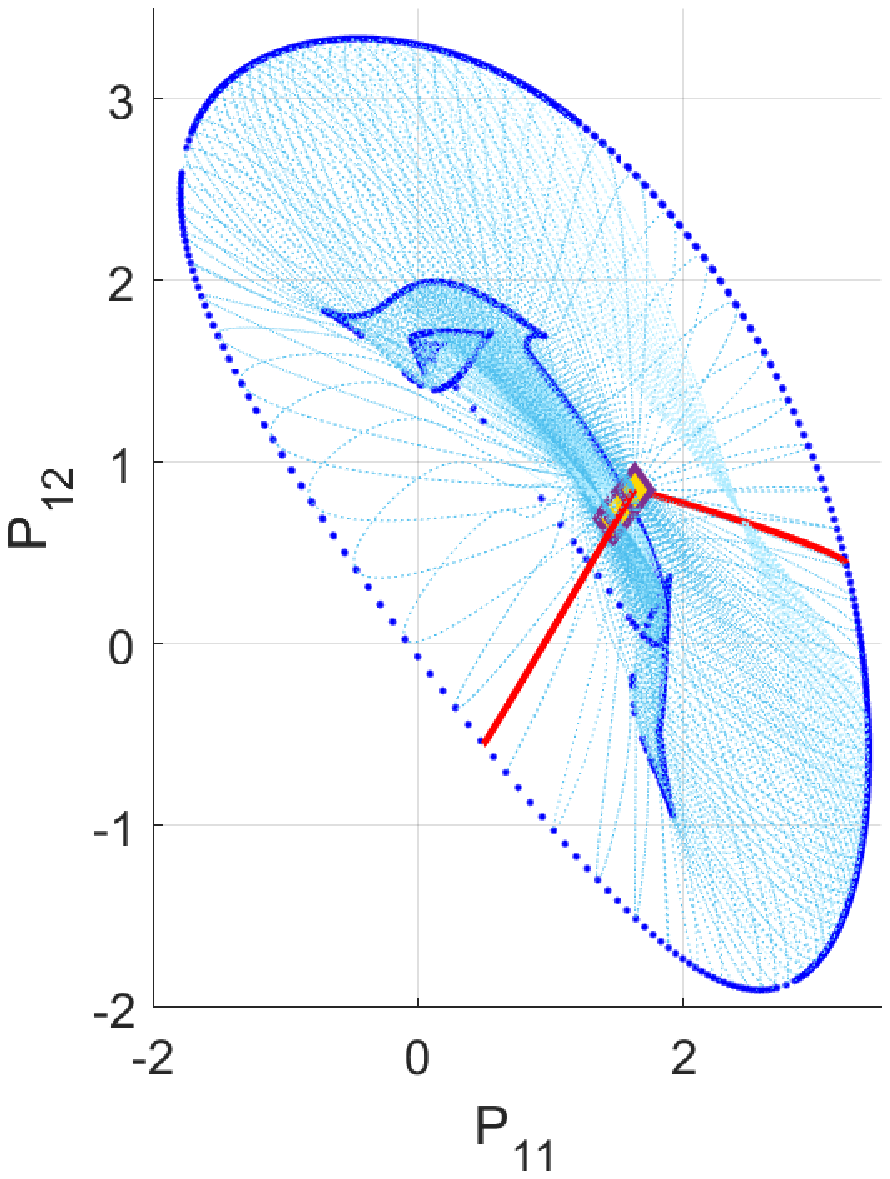}}\\
		\subfigure[Path of Euclidean Distance VS Path of Manifold Distance{\textsuperscript{c}}]{\label{fig:9_V_P_full_comp}\includegraphics[width=0.85\columnwidth]{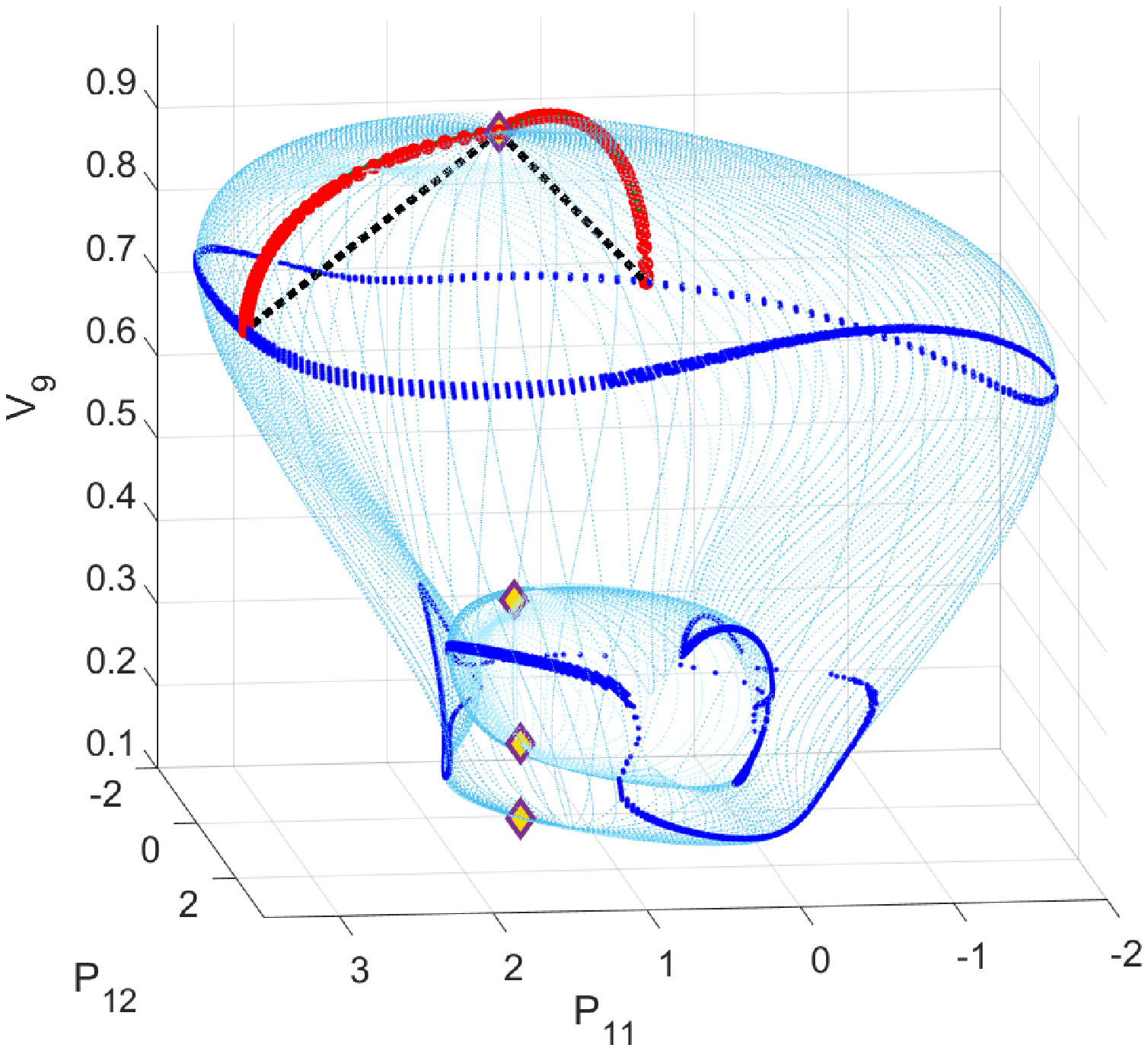}}
		\caption{Projections in Power-Voltage Subspace for Case9mod1 Static Case}
		\label{fig:9-bus_V_full_EM}
	\end{center}
	{\small\textsuperscript{a,b,c}Blue dotted curve: singular surface. \small\textsuperscript{a}Black line segment: Euclidean local min. \small\textsuperscript{b}Red curve: manifold local min. 
	\small\textsuperscript{c}Black dotted line segment: Euclidean local min, green curve: associated path for Euclidean local min, red dotted curve: manifold local min, yellow diamond: equilibrium.}
\end{figure}
\begin{figure}[tb!]
	\begin{center}
		\subfigure[Minimum Euclidean Distance{\textsuperscript{a}}]{\label{fig:9_P_P_full_Euc}\includegraphics[width=0.48\columnwidth]{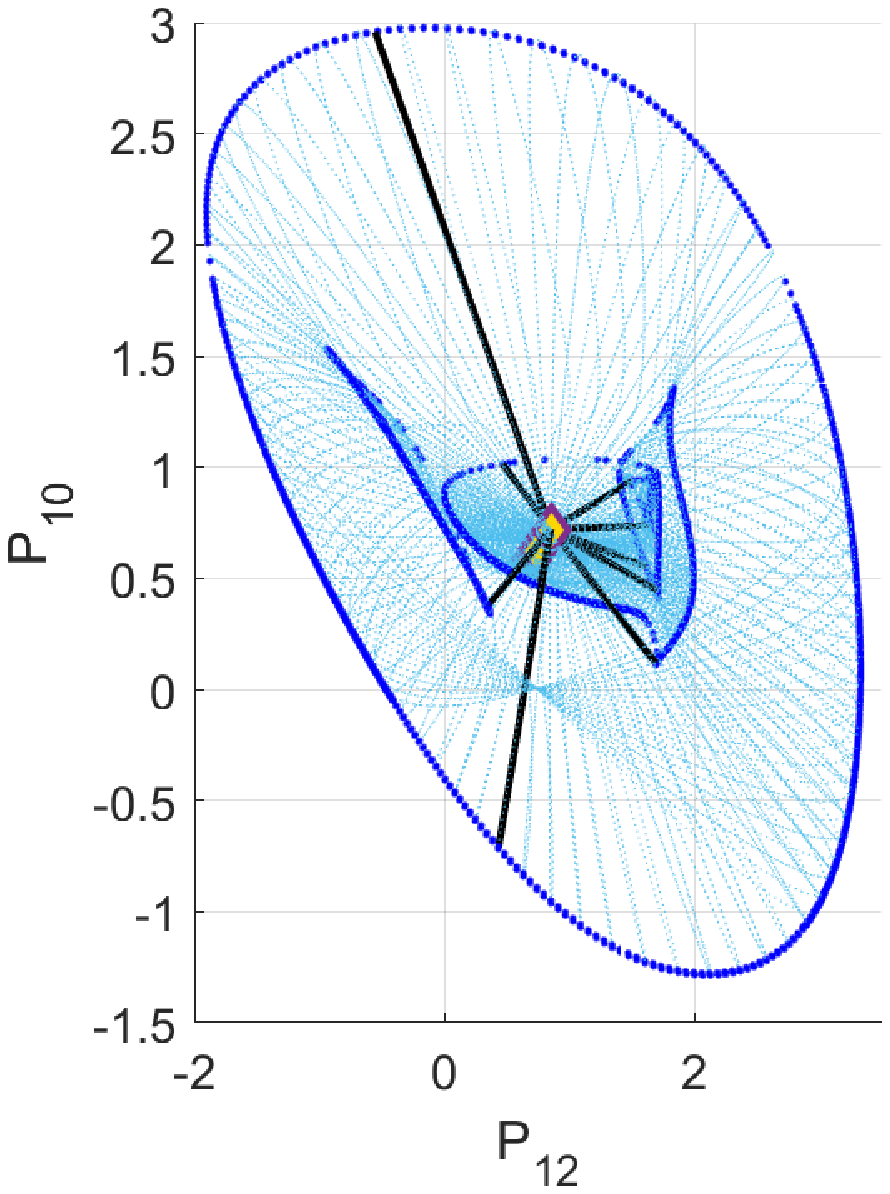}}
		\subfigure[Minimum Manifold Distance{\textsuperscript{b}}]{\label{fig:9_P_P_full_mnfld}\includegraphics[width=0.48\columnwidth]{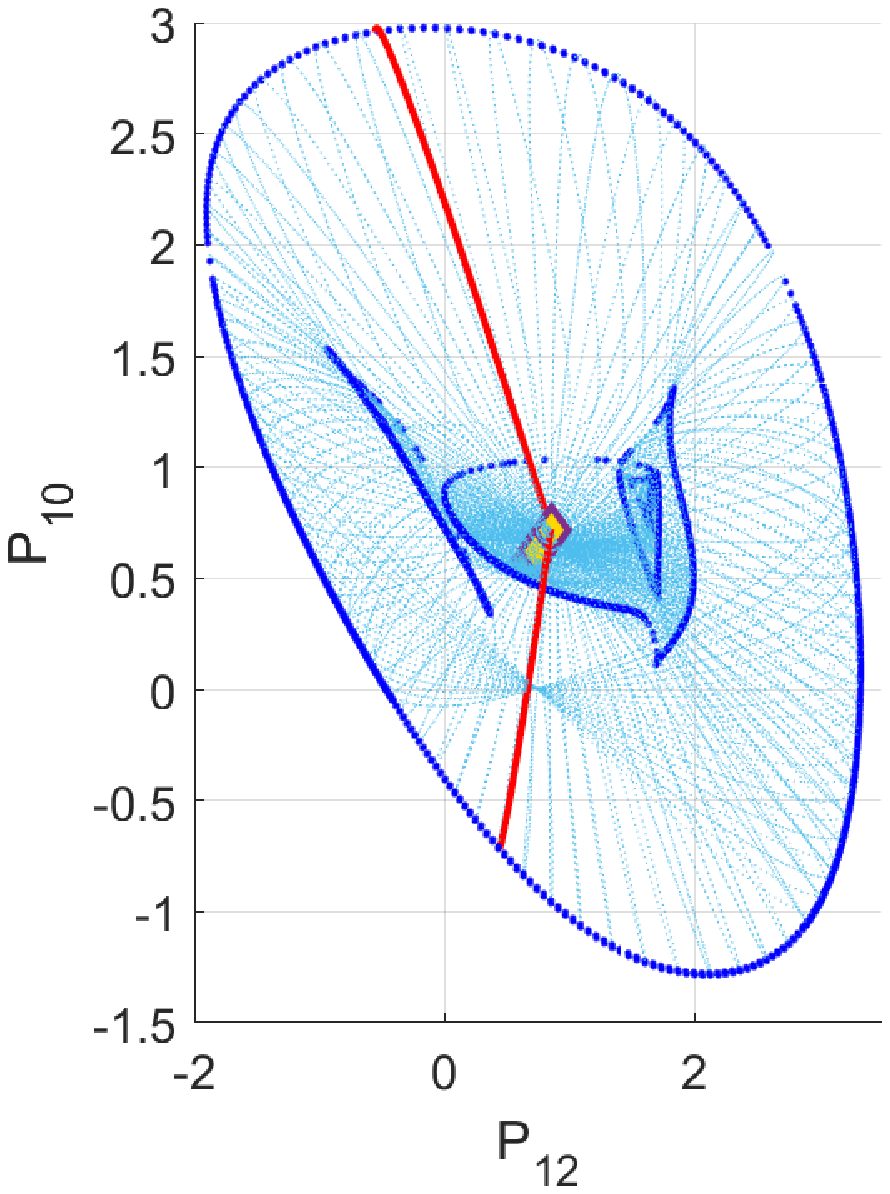}}\\
		\subfigure[Path of Euclidean Distance VS Path of Manifold Distance{\textsuperscript{c}}]{\label{fig:9_P_P_full_comp}\includegraphics[width=0.85\columnwidth]{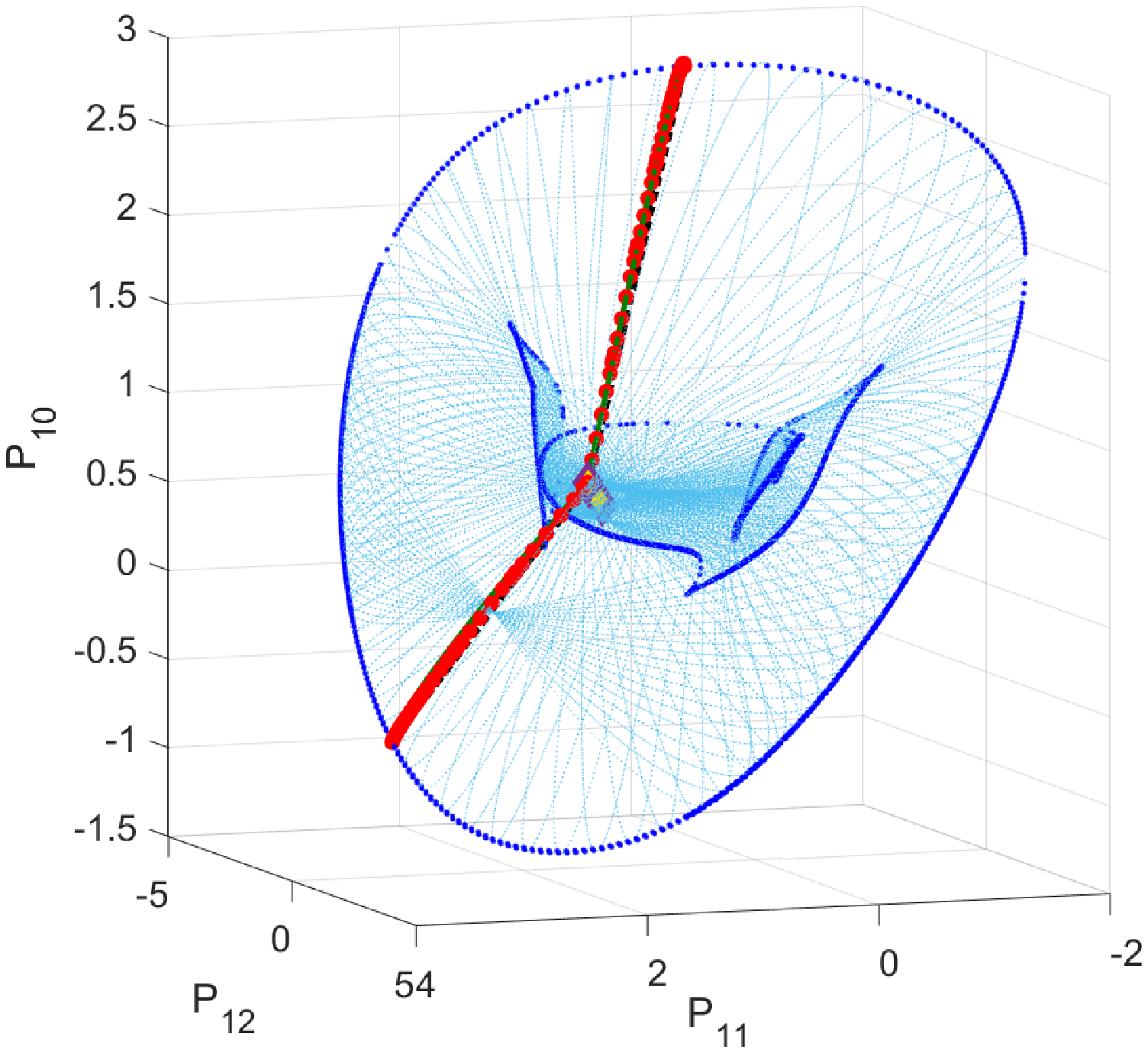}}
		\caption{Projections in Active Power Subspace for Case9mod1 Static Case}
		\label{fig:9-bus_P_full_EM}
	\end{center}
	{\small\textsuperscript{a,b,c}Blue dotted curve: singular surface. \small\textsuperscript{a}Black line segment: Euclidean local min. \small\textsuperscript{b}Red curve: manifold local min. 
	\small\textsuperscript{c}Black dotted line segment: Euclidean local min, green curve: associated path for Euclidean local min, red dotted curve: manifold local min, yellow diamond: equilibrium.}
\end{figure}

In this part, we consider the same 9-bus system ``case9mod1'' in the previous subsection, but with slow generation re-dispatch dynamics instead of fast generator swing dynamics. Such re-dispatch can be caused by the fluctuation of renewable energy  and the response of automatic generation control.  The dynamic states $x$ are the generator active power injections, while the generator angles are among the algebraic states $y$. This model takes the form of \eqref{eq:DyAE} in which case the dynamic part \eqref{eq:DyAE_Dy} needs no specification for our purpose.  
Under the constant power characteristic assumption, the singularity condition is then the commonly accepted singular power flow Jacobian matrix.

Figure~\ref{fig:9_V_P_full_comp} depicts a projection of the algebraic manifold in the Power-Voltage subspace. The static model yields two isolated singular surfaces shown by the deep blue curves. The upper one in Figure~\ref{fig:9_V_P_full_comp} is mapped to the larger outer blue curve in the active power subspace in Figure~\ref{fig:9_P_P_full_comp}. The lower one in Figure~\ref{fig:9_V_P_full_comp} is mapped to the smaller inner blue curve in Figure~\ref{fig:9_P_P_full_comp}. The formulation of Euclidean distance yields at least ten candidates of local minimum for this example, shown by the black line segments in Figure~\ref{fig:9_V_P_full_Euc} and \ref{fig:9_P_P_full_Euc}. Two of them end on the larger singular surface, and the other eight end on the smaller singular surface. The formulation of manifold distance yields two candidates of local minimal path towards voltage instability, shown in Figure~\ref{fig:9_V_P_full_mnfld} and Figure~\ref{fig:9_P_P_full_mnfld} by the red dot curves. These paths are very close to the two paths associated with the two local solutions of Euclidean distance on the larger singular surface, shown in Figure~\ref{fig:9_V_P_full_comp} and \ref{fig:9_P_P_full_comp} in green (barely distinguishable in the plots). It suggests that in this particular static example the formulation of the Euclidean distance can provide a good approximation to the distance on the manifold. A physical explanation of why the Euclidean local minimum comes close to the manifold local minimum in this model is given in Section~\ref{sec:disc} Part~D. However, as shall be seen in the following, this may not be the case when observing the distance in the reactive power subspace.

Although the formulation of Euclidean distance serves as a good approximation in this particular example, its global solution is still misleading and conservative. Figure~\ref{fig:9_V_P_full_Euc} and \ref{fig:9_P_P_full_Euc} show that the two local solutions of black line segments ending on the larger singular surface are the worst among all the ten local solutions. The other eight solutions all end on the small singular surface which cannot be reached before crossing the large singular surface first. As can be seen in Fig.~\ref{fig:9_V_P_full_comp}, to reach the lower singular surface one must cross the upper singular surface first. Moreover, since the small singular surface encloses a rather small region in the power space, the predicted stability region is also quite conservative. This can be observed from Fig.~\ref{fig:9_P_P_full_comp} that the outer singular surface encloses a much larger area than the inner singular surface.

\subsection{9-Bus Static Voltage Instability Example - Distance in Reactive Power Subspace}
\begin{figure}[tb!]
	\begin{center}
		{\includegraphics[width=0.9\columnwidth]{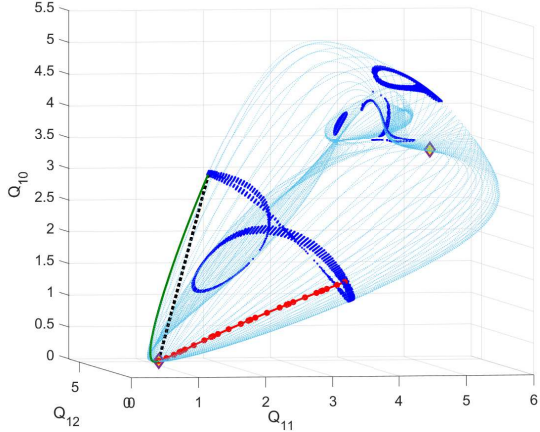}}
		\caption{Reactive Power Subspace for Case9mod2 Static Case\textsuperscript{a}}
		\label{fig:9_Qspc}
	\end{center}	
	{\small\textsuperscript{a}Blue dotted curve: singular surface, black dotted line segment: Euclidean global min, green curve: associated path for Euclidean global min, red dotted curve: manifold global min, yellow diamond: equilibrium.}
\end{figure}

Last subsection suggests that the Euclidean distance can be a good approximation for the manifold distance in the active power subspace for the static voltage stability problem. In this subsection we are going to show that the situation can be more complicated in the reactive power subspace through another modified 9-bus system example, ``case9mod2''. The parameters of the investigated example are provided in Table~\ref{table:9bus_2_bus} and \ref{table:9bus_2_branch} in Appendix~\ref{sec:Appendix I}. Our goal is to find the shortest path to the singular surface with respect to the reactive power outputs of the generators. In this particular example, we treat $30 \%$ of the load demand in Table~\ref{table:9bus_2_bus} as constant impedance and leave the rest $70 \%$ as constant power.

The manifold in the reactive power subspace is shown in Fig.~\ref{fig:9_Qspc}. The deep blue dotted curves represent isolated singular surfaces. We evaluate the local minima of Euclidean distance by initializing \eqref{eq:Euclidean} at dozens of points on the correct singular surface (the largest dotted blue curve in Fig.~\ref{fig:9_Qspc}). The primal-dual interior point solver provides four candidates of local minimum, among which the global one is shown in Fig.~\ref{fig:9_Qspc} as the black dotted straight line segment. The associated path on the manifold is depicted as the green curve with the arc length of $3.4964$. The optimal control framework, on the other hand, yields five candidates of local minimal path. We show the shortest one in Fig.~\ref{fig:9_Qspc} by dotted red curve with the arc length of $3.3048$. Although the arc length values from two formulations are not substantially different, the path directions and the ending points are totally different (see Fig.~\ref{fig:9_Qspc}) in the reactive power subspace. It suggests that in the reactive power subspace the shortest manifold distance can be completely different from the shortest Euclidean distance.

\subsection{39-Bus Static Voltage Instability Example - Distance in Complex Power Subspace}
\begin{figure}[tb!]
	\begin{center}
		{\includegraphics[width=0.9\columnwidth]{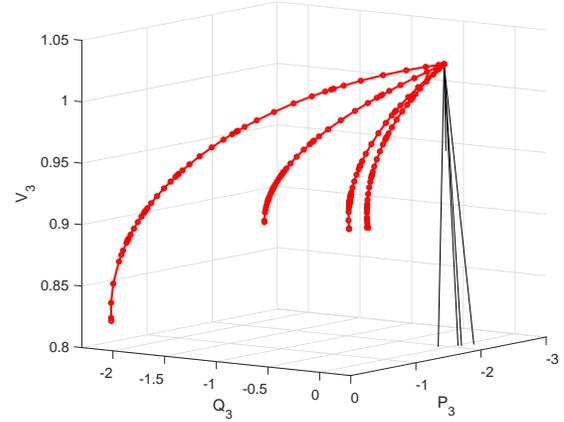}}
		\caption{Power-Voltage Subspace at Bus-3 for 39-Bus Static Case\textsuperscript{a}}
		\label{fig:39bus}
	\end{center}	
	\small\textsuperscript{a}Black line segment: Euclidean local min, red dotted curve: manifold local min.
\end{figure}
In this part we demonstrate that the proposed method can be applied to any dimensional manifold with combinations of active power and reactive power for both generator and load buses. 
Specifically, we test our proposed method on the IEEE 39-bus system. We treat $40 \%$ of the load demand as constant impedance and leave the rest $60 \%$ as constant power.

Six generator buses and six PQ buses are randomly selected as our adjustable power injection buses. In this particular study, the selected six generator buses are Bus-30, 31, 33, 34, 37 and 39. The selected six PQ buses are Bus-3, 4, 7, 8, 20 and 26. We consider the path in the subspace spanned by the selected generator active power injections and the selected load complex power demand. Therefore, the algebraic manifold $\Sigma$ is an $18$-dimensional hypersurface. 

By solving the optimal control problem \eqref{eq:oc} from different initializations we found four candidates for the local minimum paths. They are depicted in Fig.~\ref{fig:39bus} by the red dotted curves. Their arc length values are $11.9960$, $12.0729$, $16.3507$, and $16.6102$. However, after solving the Euclidean formulation \eqref{eq:Euclidean} from different initializations, $28$ candidates of local minimum are located. In Fig.~\ref{fig:39bus} five shortest ones are depicted by the black line segments. Their Euclidean distance values are $0.5821$, $1.3948$, $1.7912$, $1.8666$, and $1.8883$. Path evaluations on these solutions imply that none of them ends on the correct singular surface. If one takes the global Euclidean distance as the radius of voltage stability margin, it is at most $0.5821$, which is quite conservative comparing to the minimum arc length we found at the value of $11.9960$.

\section{Discussions}
	\label{sec:disc}

\subsection{Initialization}
Solving \eqref{eq:oc} requires an initial path on the manifold. This can be done by using the continuation power flow or the holomorphic embedding technique. In real practice, system operators usually have a prediction of how the load and renewable generation change in the next few hours. Following the predicted changing direction step by step yields a path on the manifold. This path can be used to initialize the proposed method. A local minimum path obtained from the predicted path is more informative than a local minimum path obtained from a random initialization because it characterizes the voltage stability margin in the neighborhood of the prediction.

\subsection{Voltage Stability Margin}
Problem \eqref{eq:oc} computes the (locally) shortest path to the point of voltage instability on the algebraic manifold. To have an intuitive interpretation of this path, let's first recall the traditional voltage collapse evaluation based on the continuation power flow. The continuation power flow certainly yields a path on the manifold at a fixed direction. The solution to our problem can be regarded as a direction-varying continuation power flow path which is the shortest among all possible paths (including direction-varying paths). This path is conservative but secure, especially considering the effects of uncertain nature of renewable energy in modern/future power systems.

The value of objective function \eqref{eq:oc_cost} tells the shortest manifold distance. This distance can be regarded as a stability margin (at least in the neighborhood of the path). 
A future operating point is guaranteed to be safe\footnote{The investigated operating point should not lie too far away. Otherwise another local minimum path may dominate the margin.} if its manifold distance to the current operating point is less than the margin. Alternatively, a future operating point whose distance on the manifold is greater than or close to this margin should be alarmed. 

The computation of the manifold distance to any operating point is straightforward. It only needs to remove \eqref{eq:oc_diff_r} and \eqref{eq:oc_unif} from \eqref{eq:oc} and replace the singularity condition \eqref{eq:oc_sing} by an end-point condition. Another method that can provide the shortest path between two points on the manifold is to compute the geodesic distance. A prior work on computing the geodesic distance for the power flow problem can be found in \cite{wolter:2019differential,wolter2020:differential}.
Tracking the stability margin of a varying operating point does not need to repeatedly solve the optimization problem. As long as one distance has been solved, its change can be tracked by a continuation method.

\subsection{Detecting False Voltage Stability Margin}
The numerical simulations in Section~\ref{sec:num} showed that the problem formulation of Euclidean distance can admit multiple local solutions. Many of these solutions end on the wrong singular surfaces, providing false voltage stability margins. Unfortunately, the global solution can be among them. Therefore, it raises a difficulty for the problem formulation of Euclidean distance: how can it be verified that its solution is on the correct singular surface? A prior work discussing this issue for fast-slow dynamical systems can be found in \cite{gutschke2015:differential}.

In our proposed problem formulation, it is also possible to have a solution path which ends at a wrong singular surface. However, this situation can be largely avoided and, if it happens, can be easily detected. 

Recall that we initialize problem \eqref{eq:oc} from a path given by the continuation power flow. So the initial path is usually away from the wrong singular surface. On the other hand, if the optimization solver jumps over to another singular surface, the resulting path should have more than one nose point since the path has to cross the right singular surface before reaching the wrong one. Hence, it is easy to detect the false voltage stability margin by simply counting how many nose points are on the solution path.

\subsection{Euclidean Distance VS Manifold Distance}
In Section~\ref{sec:num} Part~B we demonstrate that the Euclidean distance can be a good approximation to the manifold distance in the active power subspace for the static voltage stability problem with constant power load (generator) characteristic. Now we give an engineering explanation for this phenomenon. 

Let's denote the flexible nodal active power injection as $P_{i,t}$ and denote the fixed nodal active power injection as $P_{j,0}$. Then we have
\begin{equation}
\sum_{i} P_{i,t} + \sum_{j} P_{j,0} = P_{loss}(V) \label{eq:Ploss}
\end{equation}%
where $P_{loss}(V)$ is the active power loss of the whole system which is related to the nodal voltages. 

If the system is lossless, then $P_{loss}(V) = 0$ for any feasible $V$, which implies that \eqref{eq:Ploss} is a hyperplane with respect to $P_{i,t}$. Thus, the submanifold of active power injection $P_{i,t}$ is flat. Therefore, in the lossless case the manifold path and the Euclidean path are identical to the boundary. 

If $P_{loss}(V) \neq 0$, the submanifold of active power injection $P_{i,t}$ is no longer a hyperplane. Then, any shortest manifold path should not be an Euclidean segment generically. However, for a transmission system, $P_{loss}(V)$ is usually very small compared to the total active power injection, e.g. around $3\%$. Hence, the submanifold is very close to the flat hyperplane of the lossless case. 
However, if our load (generator) characteristic is not constant power, the singular surface in the power space can exhibit totally different structures, which thus result in very different Euclidean and manifold distances. That is exactly what happens in our 9-bus dynamic case. In Fig.~\ref{fig:9-bus_P_distP_EM} the active power subspace is also very flat and thin. But the singular surface (blue curve) occurs at the lower part of the manifold, making the green and red curves in Fig.~\ref{fig:9-bus_P_distP_EM} totally distinct from each other. A future investigation on load characteristics is promised in this framework.

On the other hand, it is common in power engineering that the transmission system usually has a large reactive power loss. Therefore, the curvature change in the reactive power subspace is more prominent than in the active power subspace, which results in a large deviation between the shortest paths of the Euclidean distance and the manifold distance in Fig.~\ref{fig:9_Qspc}. It is our ongoing research to prove the curvature properties and to evaluate the error between the manifold path and the Euclidean path given a bound on the loss function $P_{loss}(V)$. 

\subsection{Including Engineering Constraints}
In \eqref{eq:oc} we exclude any engineering constraints for simplicity. There is no modeling difficulty in adding engineering constraints, either equalities or inequalities, to \eqref{eq:oc}. For example, one can add generator ramping limits to the problem by adding bounds for the control variable in $u(\tau)$ which is associated with the adjustable generator active power output. The low voltage protection limit can be added as a lower bound for the voltage state in $x(\tau)$. Other engineering constraints are also possible. 

To solve \eqref{eq:oc} numerically, a common approach is to transcribe the differential form of \eqref{eq:oc} into a discrete form first, and then using some nonlinear optimization solver to obtain the solution for the discretized problem. Hence, adding engineering constraints does not affect the modeling and the optimization process, while it only increases the number of constraints for the transcribed problem. 

If one is interested in the voltage instability caused by the limit-induced bifurcation, then the particular engineering limit should be treated differently. Although this topic is beyond the scope of this paper, we can easily revise \eqref{eq:oc} to establish the model for solving the minimum path to the engineering limit. Suppose the condition for the limit-induced bifurcation is given by a terminal manifold
\begin{equation}
h\big( x(1), y(1), u(1), v(1)  \big)=0 \label{eq:end_mnfld}
\end{equation}%
Then, we remove the auxiliary state vector $r(\tau)$ and its associated conditions \eqref{eq:oc_diff_r} and \eqref{eq:oc_unif} from \eqref{eq:oc}, and replace \eqref{eq:oc_sing} by \eqref{eq:end_mnfld}.

\section{Conclusion} 
   	\label{sec:concl}
   	This paper discusses how to find the (locally) shortest path to the point of voltage collapse. We first recall different mechanisms of voltage collapse. Then, we focus on a particular class of voltage instability which is induced by the singularity condition of the algebraic manifold. Instead of identifying the shortest path with the Euclidean distance, we establish the distance on the manifold and formulate a general optimization framework that can incorporate the manifold distance. To solve this optimization problem, we further convert it into an optimal control framework and solve it through a standard optimal control solver.

The proposed problem formulation is more rigorous and meaningful than the formulation of Euclidean distance because it respects manifold curvature change for the entire path. Numerical simulations validate the proposed approach by comparing it to the solutions obtained from the formulation of Euclidean distance. We specifically demonstrate that the shortest Euclidean distance may not be the shortest distance on the manifold, and can be misleading and conservative when it ends on a wrong singular surface. Unlike the potential issues with conventional approaches, the proposed approach yields the correct path on the manifold for all the tested cases. 

A promising future research direction can be finding an operating point away from voltage collapse in the proposed framework. It allows the design of voltage control that respects the manifold. 

\section*{Acknowledgement} 
The authors gratefully acknowledge the support from NSF Grand CNS 1735463. We would like to thank the helpful discussions with Sam Chevalier, Prof. Eytan Modiano, Dr. Marija Ilic, Dr. Xia Miao and Dr. Xinbo Geng. 

\bibliographystyle{ieeetr}
\bibliography{References}

\appendices
\section{9-Bus Systems Parameters} 
\label{sec:Appendix I}
\begin{table}[!htbp]
	\centering
\begin{threeparttable}
	\caption{Case9mod1 Node Parameters}
	\begin{tabular}{c|c|c|c|c}
		\hline
		Bus \# & Type & P (MW) & Q (MVar) & Voltage    \\ \hline
		1   & 1    & 0   &  0  &        \\
		2   & 1    & 0   &  0  &        \\
		3   & 1    & 0   &  0  &        \\
		4   & 1    & 0   &  0  &        \\
		5   & 1    & -90 & -30  &        \\
		6   & 1    & 0   &  0  &        \\
		7   & 1    & -100& -35 &        \\
		8   & 1    & 0   &  0  &        \\
		9   & 1    & -125& -50 &        \\ \hline
		10  & 3    &     &      & 1.0388 \\
		11  & 2    & 163.1587  &     & 1.0264 \\
		12  & 2    & 85.0429   &    & 1.0003 \\ \hline
	\end{tabular}\label{table:9bus_1_bus}
	\begin{tablenotes}[flushleft]
		\small
		\item Type 1 is the PQ bus. Type 2 is the PV bus. Type 3 is the slack bus.
		\item Base power is $100$ MVA.
	\end{tablenotes}
\end{threeparttable}
\end{table}

\begin{table}[!htbp]
	\centering
	\begin{threeparttable}
		\caption{Case9mod1 Branch Parameters}
		\begin{tabular}{c|c|c|c|c}
			\hline
			From Bus & To Bus & r                       & x      & b      \\ \hline
			1        & 4      & $10^{-5}$               & 0.0576 & 0      \\
			4        & 5      & 0.0170                  & 0.0920 & 0.1580 \\
			5        & 6      & 0.0390                  & 0.1700 & 0.3580 \\
			3        & 6      & $10^{-5}$               & 0.0586 & 0      \\
			6        & 7      & 0.0119                  & 0.1008 & 0.2090 \\
			7        & 8      & 0.0085                  & 0.0720 & 0.1490 \\
			8        & 2      & $10^{-5}$               & 0.0625 & 0      \\
			8        & 9      & 0.0320                  & 0.1610 & 0.3060 \\
			9        & 4      & 0.0100                  & 0.0850 & 0.1760 \\
			10       & 1      & $6.8670 \times 10^{-4}$ & 0.1391 & 0      \\
			11       & 2      & $5.9259 \times 10^{-4}$ & 0.0948 & 0      \\
			12       & 3      & $5.9259 \times 10^{-4}$ & 0.0948 & 0     \\ \hline
		\end{tabular}\label{table:9bus_1_branch}
		\begin{tablenotes}[flushleft]
			\small
			\item r, x and b are in per unit.
		\end{tablenotes}
	\end{threeparttable}
\end{table}

\begin{table}[!htbp]
	\centering
	\begin{threeparttable}
		\caption{Case9mod2 Node Parameters}
		\begin{tabular}{c|c|c|c|c}
			\hline
			Bus \# & Type & P (MW) & Q (MVar) & Voltage    \\ \hline
			1   & 1    & 0   &  0  &        \\
			2   & 1    & 0   &  0  &        \\
			3   & 1    & 0   &  0  &        \\
			4   & 1    & 0   &  0  &        \\
			5   & 1    & -90 & -50  &        \\
			6   & 1    & 0   &  0  &        \\
			7   & 1    & -100& -50 &        \\
			8   & 1    & 0   &  0  &        \\
			9   & 1    & -125& -50 &        \\ \hline
			10  & 3    &     &      & 1.0331 \\
			11  & 2    & 150.1369  &     & 1.0340 \\
			12  & 2    & 150.1351   &    & 1.0274 \\ \hline
		\end{tabular}\label{table:9bus_2_bus}
		\begin{tablenotes}[flushleft]
			\small
			\item Type 1 is the PQ bus. Type 2 is the PV bus. Type 3 is the slack bus.
			\item Base power is $100$ MVA.
		\end{tablenotes}
	\end{threeparttable}
\end{table}

\begin{table}[!htbp]
	\centering
	\begin{threeparttable}
		\caption{Case9mod2 Branch Parameters}
		\begin{tabular}{c|c|c|c|c}
			\hline
			From Bus & To Bus & r                       & x      & b      \\ \hline
			1        & 4      & 0.0010               & 0.0576 & 0      \\
			4        & 5      & 0.0170                  & 0.0920 & 0.1580 \\
			5        & 6      & 0.0190                  & 0.0600 & 0.3580 \\
			3        & 6      & 0.0010               & 0.0586 & 0      \\
			6        & 7      & 0.0119                  & 0.0608 & 0.2090 \\
			7        & 8      & 0.0085                  & 0.0620 & 0.1490 \\
			8        & 2      & 0.0010               & 0.0625 & 0      \\
			8        & 9      & 0.0120                  & 0.0610 & 0.3060 \\
			9        & 4      & 0.0100                  & 0.0850 & 0.1760 \\
			10       & 1      & $6.8670 \times 10^{-4}$ & 0.1391 & 0      \\
			11       & 2      & $5.9259 \times 10^{-4}$ & 0.0948 & 0      \\
			12       & 3      & $5.9259 \times 10^{-4}$ & 0.0948 & 0     \\ \hline
		\end{tabular}\label{table:9bus_2_branch}
		\begin{tablenotes}[flushleft]
			\small
			\item r, x and b are in per unit.
		\end{tablenotes}
	\end{threeparttable}
\end{table}

\section{Detailed Optimal Control Formulation for Static Voltage Problem in Active Power Generation Subspace}
\label{sec:Appendix II}
Consider a power grid with $N_{bus}$ many nodes, among which $N_{gen}$ is the number of generator buses and $N_{load}$ is the number of load buses. If we are interested in the path in the active power generation subspace (as the second example in Section~\ref{sec:num}), the following formulation is applied. 
\begin{subequations}
	\begin{align}	
	\mbox{\bf Minimize}~~ & \int_0^1 \sqrt{\langle u(\tau), u(\tau)\rangle} d\tau \label{eq:9bus_cost} \\ 
	\mbox {\bf Subject to:}~~ & \frac{d}{d\tau} P_{gen}(\tau) = u(\tau) \label{eq:9bus_dpgen}  \\ 
	& \frac{d}{d\tau} V(\tau) = v(\tau) \label{eq:9bus_dV}  \\
	& \frac{d}{d\tau} r(\tau) = \mathbf{0} \label{eq:9bus_dr}  \\
	& P_{gen,i}(\tau)-f_{gen,i}\big( V(\tau) \big) = 0 \label{eq:9bus_pgen}  \\
	& h_{gen,i}\big( V(\tau) \big) = 0 \label{eq:9bus_pv}  \\
	& f_{load,j}\big( V(\tau) \big) = 0 \label{eq:9bus_pload}  \\
	& g_{load,j}\big( V(\tau) \big) = 0 \label{eq:9bus_qload}  \\
	& \left[ \begin{array}{c}
	\partial f_{gen}(V)/\partial V \\
	\partial h_{gen}(V)/\partial V \\
	\partial f_{load}(V)/\partial V \\
	\partial g_{load}(V)/\partial V
	\end{array} \right]_{\tau=1}^T r(1) = \mathbf{0} \label{eq:9bus_sing} \\
	& \langle r(1), r(1) \rangle = 1 \label{eq:9bus_unit} \\
	& P_{gen}(0) = P_{gen}^0 \label{eq:9bus_pgen0} \\
	& V(0) = V^0 \label{eq:9bus_V0} \\
	\mbox{Index:}~~ & i=1, \dots, N_{gen}\nonumber\\
	& j=N_{gen}+1, \dots, N_{bus}	\nonumber%
	\end{align}\label{eq:acopf}%
\end{subequations}%
where $P_{gen} \in \Rc^{N_{gen}}$ is the generator active power injection vector that serves as the dynamic state vector; $u$ is the control vector associated with $P_{gen}$; $V \in \Rc^{2 N_{bus}-1}$ is the node voltage vector (algebraic state vector) in rectangular coordinates without the angle reference element; $v$ is the control vector associated with $V$; $r \in \Rc^{2 N_{bus}-1}$ is the auxiliary state vector to enforce singularity condition; $f_{gen,i}$ is the network active power function at generator node-$i$; $h_{gen,i}$ is the node voltage magnitude function at generator node-$i$; $f_{load,j}$ is the network active power function at load node-$j$; $g_{load,j}$ is the network reactive power function at load node-$j$. The Jacobian matrix in Eqt~\eqref{eq:9bus_sing} should exclude the active power components at the slack node. 

\end{document}